\documentclass[pra,aps,superscriptaddress,twocolumn,floatfix,nofootinbib]{revtex4-1}

\usepackage{dsfont}
\usepackage{mathtext}
\usepackage{mathtools}
\usepackage[usenames,dvipsnames]{xcolor}
\usepackage{amsmath}
\usepackage{amssymb,mathrsfs}
\usepackage{graphicx}
\usepackage{epstopdf}
\usepackage[colorlinks=true]{hyperref}
\usepackage{mathtext}

\makeatletter
\renewcommand\@make@capt@title[2]{%
    \@ifx@empty\float@link{\@firstofone}{\expandafter\href\expandafter{\float@link}}%
        {\textbf{#1}}\@caption@fignum@sep#2\quad}%
\makeatother

\makeatletter
\renewcommand{\fnum@figure}{\textbf{Fig.~\thefigure}}
\makeatother

\newcommand{\al}{\alpha}
\newcommand{\om}{\omega}

\newcommand{\tk}{\widetilde{k}}
\newcommand{\tom}{\widetilde{\omega}}
\newcommand{\tal}{\widetilde{\alpha}}
\newcommand{\twb}{t_{\rm\scriptscriptstyle WB}}

\newcommand{\prt}{\partial}

\newcommand {\ox}{\overline{x}}
\newcommand {\orho}{\overline{\rho}}
\newcommand {\ovu}{\overline{u}}

\newcommand{\rom}[1]{\uppercase\expandafter{\romannumeral #1\relax}}

\newcommand{\sss}{\scriptscriptstyle}

\makeatletter

\renewcommand*{\p@subsection}{}

\renewcommand*{\p@subsubsection}{}
\makeatother

\graphicspath{{Figures/}}

\begin{document}

\title{Formation of dispersive shock waves in a saturable nonlinear medium}

\author{Sergey~K.~Ivanov} \affiliation{Moscow Institute of Physics and
  Technology, Institutsky lane 9, Dolgoprudny, Moscow region, 141700,
  Russia} \affiliation{Institute of Spectroscopy, Russian Academy of
  Sciences, Troitsk, Moscow, 108840, Russia}

\author{Jules-El\'{e}mir~Suchorski} \affiliation{Universit\'e
  Paris-Saclay, CNRS, LPTMS, 91405, Orsay, France}

\author{Anatoly~M.~Kamchatnov} \affiliation{Moscow Institute of
  Physics and Technology, Institutsky lane 9, Dolgoprudny, Moscow
  region, 141700, Russia} \affiliation{Institute of Spectroscopy,
  Russian Academy of Sciences, Troitsk, Moscow, 108840, Russia}

\author{Mathieu~Isoard} \affiliation{Universit\'e Paris-Saclay, CNRS,
  LPTMS, 91405, Orsay, France}

\author{Nicolas~Pavloff} \affiliation{Universit\'e Paris-Saclay, CNRS,
  LPTMS, 91405, Orsay, France}

\begin{abstract}
  We use the Gurevich-Pitaevskii approach based on the Whitham
  averaging method for studying the formation of dispersive shock
  waves in an intense light pulse propagating through a saturable
  nonlinear medium. Although the Whitham modulation equations cannot
  be diagonalized in this case, the main characteristics of the
  dispersive shock can be derived by means of an analysis of the
  properties of these equations at the boundaries of the
  shock. Our approach generalizes a previous analysis
    of step-like initial intensity distributions to a more realistic
    type of initial light pulse and makes it possible to determine, in
    a setting of experimental interest, the value of measurable
    quantities such as the wave-breaking time or the position and
    light intensity of the shock edges.
\end{abstract}




\maketitle

\section{Introduction} \label{sec1}

The propagation of nonlinear waves in dispersive media has attracted
much attention in various fields of research such as water waves,
plasma physics, nonlinear optics, Bose-Einstein condensates and
others. In particular, the expansion of an initial state with a fairly
smooth and large profile is accompanied by gradual steepening followed
by wave breaking resulting in the formation of a dispersive shock wave
(DSW), and this phenomenon was experimentally observed, for example,
in Bose-Einstein condensates \cite{Dut01,secscb-05,hacces-06} and nonlinear
optics \cite{gcrt-07,ggfdc-12,wmktcc-16,Xu2017,mbgc-17,nfxmef-19}.  In these
systems, the dynamics of the pulse is well described by the nonlinear
Schr\"odinger (NLS) or Gross-Pitaevskii equation and the initial stage
of evolution admits a purely hydrodynamic description in terms of the
classical Riemann method, see, e.g., Refs.
\cite{For2009,ik-19,ikp-19a,ikp-19b,ikp-19c}.  After the wave breaking
moment, when DSWs are formed, the evolution of the wave structure can
be described by the Whitham method
\cite{Whitham-65,Whitham-74,kamch-2000,eh-16}. In the case of cubic
Kerr-like nonlinearity the NLS equation is completely integrable,
hence the Whitham modulation equations can be put in a diagonal
Riemann form \cite{fl86,pavlov87}, and, with the use of the
Gurevich-Pitaevskii approach \cite{gp-73}, it has been possible to
develop a detailed analytic theory of the evolution of DSWs
\cite{gk-87,eggk-95,ek-95,ikp-19a}. This produces an excellent
description of experiments \cite{wmktcc-16,Xu2017} on evolution of initial
discontinuities in the intensity distribution of optical pulses propagating
in optical fibers.

As first understood by Sagdeev~\cite{Moi63} in the context of viscous
stationary shocks whose width is much greater that a typical soliton
width, the formation of DSWs is a universal phenomenon which occurs in
a number of systems demonstrating nonlinear dispersive waves. In
particular, the formation of DSWs has been observed in the propagation
of light beams through a photorefractive medium
\cite{cmc-04,wjf-07}. However, in such a system the nonlinearity is
not of the Kerr type, and the theory of
Refs.~\cite{gk-87,eggk-95,ek-95,ikp-19a} thus needs to be modified.
Of course, the validity of the Whitham averaging approach for the
description of the modulations of a nonlinear wave does not depend on
the complete integrability of the wave equation. Nevertheless, it is
difficult to put it into practice for non-completely integrable
equations because of the lack of diagonalized form of the Whitham
modulation equations.

The first general statement about the properties of DSWs applicable to
a non-integrable situation was made by Gurevich and Meshcherkin in
Ref.~\cite{gm-84}.  In this work the authors claimed that, when a DSW
is formed after wave breaking of a ``simple wave'', that is of a wave
for which one of the non-dispersive Riemann invariants is constant,
then the value of this constant Riemann invariant remains the same at
both extremities of the DSW.  In other words, this means that such a
wave breaking leads to the formation of a single DSW, at variance with
the situation with viscous shocks where more complicated wave
structures are generated (see, e.g., Ref. \cite{LL-6}).  The next
important step was made by El \cite{El-05} who showed that, in
situations of the Gurevich-Meshcherkin type, when the Whitham
equations at the small-amplitude edge include the linear ``number of
waves'' conservation law, this equation can be reduced to an ordinary
differential equation whose solution provides the wave number of a
linear wave at this edge. Under some reservations, a similar approach
can be developed for the soliton edge of the DSW.  As a result, one
can calculate the speeds of both edges (solitonic, large amplitude
one, and linear, small amplitude one) in the important case of a
step-like initial condition. This approach was applied to many
concrete physical situations
\cite{El-05,ElGrimshaw-06,egkkk-07,ElGrimshaw-09,EslerPearce-11,Hoefer-14,CKP-16,HEK-17,AnMarchant-18},
and was recently extended in Ref. \cite{Kamchatnov-19} to a general
case of a simple-wave breaking and then applied to shallow water waves
described by the Serre equation \cite{IvanovKamchatnov-2019}. In the
present paper, we use the same approach to study the propagation of
optical pulses and beams through a saturable nonlinear medium. This
makes it possible to considerably extends the theory of
Ref.~\cite{egkkk-07} and provides a more realistic explanation of the
results of Ref.~\cite{wjf-07} and of future experimental studies in
nonlinear optics \cite{Bienaime}.

\section{Formulation of the problem}\label{sec2}

The formation of DSWs has been observed in the spatial evolution of
light beams propagating through self-defocusing photorefractive
crystals in Refs.~\cite{cmc-04,wjf-07}. Initial non-uniformities of
the beam give rise to breaking singularities resulting in the
formation of dispersive shocks. As is well known \cite{LL-8}, in the
paraxial approximation, the propagation of the complex amplitude
$A=A(X,Y,Z)$ of the electric field of a monochromatic beam is
described by the equation
\begin{equation} \label{NLS2D}
\begin{split}
    i \frac{\partial A}{\partial Z} +
\frac{1}{2k_0n_0}\Delta_\perp A + k_0 \delta n(I)A = 0,
\end{split}
\end{equation}
where $k_0=2\pi/\lambda$ is the carrier wave number, $Z$ is the
coordinate along the beam, $X$, $Y$ are transverse coordinates,
$\Delta_\perp=\partial^2/\partial^2X+\partial^2/\partial^2Y$ is the
transverse Laplacian, $n_0$ is the linear refractive index, and
$\delta n$ is a nonlinear index which depends on the light intensity
$I=|A|^2$. It is often the case that the nonlinearity is not of pure
Kerr type (i.e., not exactly proportional to $I$) but saturates at
large intensity. We consider here the case of a defocusing saturable
medium where $\delta n$ is of the form
\begin{equation} \label{PhotorefNolin}
\begin{split}
    \delta n=-n_2\, \frac{I}{I+I_{\rm sat}},
\end{split}
\end{equation}
where $n_2$ is a constant positive coefficient. This situation is
encountered for instance in semiconductor doped glasses \cite{Cou91}
and in photorefractive media \cite{Kiv2003}. In this
latter case $n_2$ linearly depends on the applied electric field and
on the electro-optical index.  A near-resonant laser field propagating
inside a hot atomic vapor is also described by a saturable mixed
absorptive and dispersive susceptibility \cite{Boy92,Wan04}. At
negative detuning the nonlinearity is defocusing, and if the detuning
is large enough, absorption effects are small compared to nonlinear
ones. As a result, the propagation of the beam is described by an
envelope equation which can be cast in the form of Eq.~\eqref{NLS2D},
with a nonlinearity of the type \eqref{PhotorefNolin}, see e.g.,
Ref. \cite{San18,Fon18}.

We define dimensionless units by choosing a reference intensity
$I_{\rm ref}$ which can be chosen for instance as the background
intensity in the situation considered in Refs.~\cite{cmc-04,wjf-07}
where the $Z=0$ light distribution has the form of a region of
decreased \cite{cmc-04} or increased \cite{wjf-07} light intensity
perturbating a uniform background. Another natural choice would be
$I_{\rm ref}=I_{\rm sat}$. We define dimensionless variables
\begin{equation} \label{NondimVar}
\begin{split}
    t=&k_0 n_2 \, \frac{I_{\rm ref}}{I_{\rm sat}}\, Z,   \quad
    x=k_0\sqrt{n_0\, n_2\, \frac{I_{\rm ref}}{I_{\rm sat}}}\, X, \\
    y=&k_0\sqrt{n_0\, n_2\,\frac{I_{\rm ref}}{I_{\rm sat}}}\, Y,  \quad
\psi = \frac{A}{\sqrt{I_{\rm ref}}}.
\end{split}
\end{equation}
We consider a geometry where the transverse profile is translationally
invariant and depends on a single coordinate $x$. Then, the
dimensionless generalized NLS equation \eqref{NLS2D} takes the form
\begin{equation} \label{gNLS}
\begin{split}
  i \psi_t + \frac{1}{2}\psi_{xx} -
  \frac{|\psi|^2}{1+\gamma|\psi|^2} \psi = 0,
\end{split}
\end{equation}
where $\gamma = I_{\rm ref}/I_{\rm sat}$. When the light intensity is small
compared to the saturation intensity, i.e., when
$\gamma|\psi|^2\ll 1$, Eq.~\eqref{gNLS} reduces to the usual
defocusing NLS equation; but at large intensity the nonlinearity
saturates.

The transition from the function $\psi$ to the dimensionless intensity
$\rho$ and chirp $u$, is performed by means of the Madelung
transform
\begin{equation} \label{Madelung}
    \psi(x,t) = \sqrt{\rho(x,t)}\,\exp{\left( i \int^x u(x',t)dx'\right)}.
\end{equation}
After substitution of this expression into Eq.~(\ref{gNLS}),
separation of the real and imaginary parts
and differentiation of one of the equations with respect to $x$,
we get the system
\begin{equation} \label{Hydro}
\begin{split}
    & \rho_t+(\rho u)_x=0, \\
    & u_t + uu_x + \frac{\rho_x}{(1+\gamma\rho)^2} +
    \left(\frac{\rho^2_x}{8\rho^2}-\frac{\rho_{xx}}{4\rho}\right)_x = 0.
\end{split}
\end{equation}
In the problem of light beam propagation,
the last term of the left-hand side of the second equation accounts
for diffractive effects. It is sometimes referred to as a
``dispersive term'', or ``quantum pressure'' or ``quantum potential''
depending on the context. The Madelung transform reveals that light
propagating in a nonlinear medium behaves as a fluid, amenable to a
hydrodynamic treatment, see Ref. \cite{ask-67}. The light
intensity $\rho$ in the hydrodynamic formulation \eqref{Hydro} is
interpreted as the density of this effective fluid and the chirp $u$
is the effective flow velocity. The coordinate $t$ along the beam
plays the role of time, and in this case the last term in
(\ref{Hydro}) should be referred to as ``dispersive''.

The present paper in organized in two main parts. In the first one
(Sec.~\ref{sec3}) we describe the dispersionless evolution of a light
pulse in the presence of a background. After a certain distance, the
wave breaks and a DSW is formed. The main characteristics of this
shock wave are studied in the second part of the work
(Sec.~\ref{sec4}).

We consider the case where the initial profile has the form of an
increased parabolic intensity bump over a constant and stationary
background (with $u(x,t=0)=0$).  If this profile is smooth enough (in
a way which will be quantified in Sec.~\ref{sec22}) the separation of
the initial bump in two counter propagating pulses can be described by
a dispersionless approach: this is performed in Sec.~\ref{sec3}.  The
difficulty lies in the fact that the initial profile is not a simple
wave in a hydrodynamic sense of this terminology.
This means that the regions where the initial non-dispersive
Riemann invariants depend on position significantly
overlap. Therefore, to study this problem, we should resort to the
Riemann hodograph method \cite{For2009,ik-19,ikp-19a,ikp-19b,ikp-19c}
which we specialize for the current photorefractive system in
Sec.~\ref{sec21}.  Since the initial intensity profile has a
discontinuity in its first derivative, then in the process of
evolution, two simple rarefaction waves are formed along the edges
(they are described in Sec.~\ref{sec23}), and the central part is the
region where the hodograph transform should be employed.  Due to
nonlinearity, the profiles at both extremities of the pulse gradually
steepen and this leads to wave breaking after some finite sample
length, or equivalently some finite ``time'' which we denote as the
wave breaking time, $\twb$. Typically this occurs in a
region where only one Riemann invariant varies, and the corresponding
DSW results from the breaking of a simple wave, which is the case
considered in the second part of the present paper.

After the wave breaking, the dispersive effects
cannot longer be neglected. For their description, we resort in
Sec.~\ref{sec4} to Whitham modulation theory
\cite{Whitham-65,Whitham-74} which is based on the large difference in
spatial and temporal scales between the rapid nonlinear oscillations
and their slow envelope.
However, Eq.~\eqref{gNLS} being non-integrable, the Whitham modulation
equations cannot be transformed to a diagonal Riemann form: the lack
of Riemann invariants hinders the application of the full Whitham
modulation theory to our system after wave breaking.  Nonetheless, the
method of Refs.~\cite{El-05,Kamchatnov-19} is based on the universal
applicability of Whitham's ``number of waves conservation law'' as
well as on the conjecture of the applicability of its soliton
counterpart to the above mentioned class of initial conditions. Such
an approach is substantiated by comparison with similar situations in
the case of completely integrable wave equations. It makes it possible
to calculate the limiting characteristic velocities of the Whitham
modulation equations at the boundary with the smooth part of the pulse
whose evolution obeys the dispersionless approximation equations, even
after the wave breaking time $\twb$. We will treat
in two separate subsections the case of a positive (Sec.~\ref{sec41})
and of a negative (Sec.~\ref{sec42}) initial intensity pulse.

Having formulated the problem, we now proceed to
the study of the dispersionless stage of evolution.

\section{Dispersionless Stage}\label{sec3}

For smooth enough wave patterns we can neglect the dispersive
term in the second equation of the system (\ref{Hydro}) and the
initial evolution of the system is described by the so-called
dispersionless equations which can be put in a form equivalent to the
equations of inviscid gas dynamics
\begin{equation} \label{HydroDisp}
\rho_t+(\rho u)_x=0, \quad u_t + uu_x + \frac{c^2(\rho)}{\rho} \rho_x  = 0,
\end{equation}
where
\begin{equation} \label{LocalVel}
c(\rho) = \frac{\sqrt{\rho}}{1+\gamma\rho}
\end{equation}
is the local sound velocity in the medium.
These equations can be cast into a diagonal form by
introducing new variables, known as the Riemann invariants
\begin{equation} \label{RiemannInv}
\begin{split}
  r_{\pm}(x,t)= &
  \frac{u(x,t)}{2}\pm \frac{1}{2}
\int_0^{\rho(x,t)} \frac{c(\rho')}{\rho'}\,d\rho'\\
= &   \frac{u(x,t)}{2}\pm
  \frac{1}{\sqrt{\gamma}}\arctan{\sqrt{\gamma\rho(x,t)}},
\end{split}
\end{equation}
whose evolution is described by the following equations
\begin{subequations}\label{RiemannEquat}
\begin{align}
\label{RiemannEquat+}
\frac{\partial r_{+}}{\partial t}
+v_{+}(r_{+},r_{-})\frac{\partial r_{+}}{\partial x}=0,\\
\label{RiemannEquat-}
\frac{\partial r_{-}}{\partial t}
+v_{-}(r_{+},r_{-})\frac{\partial r_{-}}{\partial x}=0,
\end{align}
\end{subequations}
with Riemann velocities
\begin{equation} \label{RiemannVel}
v_{\pm}=u\pm c.
\end{equation}
In this last equation, the dependence of $u$ and $c$ on $r_+$ and
$r_-$ is obtained from Eqs.~\eqref{RiemannInv}. This yields
\begin{equation}
  u = r_++r_-, \label{PhysVar_u}
\end{equation}
whereas $c$ is computed as follows: one inverts the relation
\begin{equation}\label{crprm}
r_+-r_-=\int_0^{\rho} \frac{c(\rho')}{\rho'}\,d\rho',
\end{equation}
which yields
\begin{equation}
\rho = \frac{1}{\gamma}
\tan^2{\left[\frac{\sqrt{\gamma}}{2}(r_+-r_-) \right] },\label{PhysVar_rho}
\end{equation}
and then one evaluates $c(\rho(r_+-r_-))$ from \eqref{LocalVel}.
Once the solution of Eqs.~(\ref{RiemannEquat}) are found and
the Riemann invariants are known, then the physical variables
are obtained from relations \eqref{PhysVar_u} and \eqref{PhysVar_rho}.

\subsection{Riemann method}\label{sec21}

The Riemann method which we presenting now, enables one to find the
solutions of Eqs.~(\ref{RiemannEquat}) in the generic region where
both Riemann invariants are changing.  Riemann noticed that
Eqs.~(\ref{RiemannEquat}) become linear with respect to the
independent variables $x$ and $t$ if these are considered as functions
of the Riemann invariants: $x = x(r_+, r_-)$, $t = t(r_+, r_-)$. After
this ``hodograph transform'' from the real space $(x,t)$ to the
hodograph plane $(r_+,r_-)$ we arrive at the system
\begin{equation}\label{HodTransEq}
\begin{split}
\frac{\partial x}{\partial r_-}-
v_{+}(r_{+},r_{-})\frac{\partial t}{\partial r_-} & = 0, \\
\frac{\partial x}{\partial r_+}-
v_{-}(r_{+},r_{-})\frac{\partial t}{\partial r_+} & = 0.
\end{split}
\end{equation}
It should be noticed that the Jacobian of this transformation is equal to
\begin{equation}\label{}
\begin{split}
J=\left| \frac{\partial (x,t)}{\partial (r_+,r_-)} \right| =
\frac{\partial t}{\partial r_+}\frac{\partial t}{\partial r_-} (v_--v_+),
\end{split}
\end{equation}
thus the hodograph transform breaks down whenever
${\partial t}/{\partial r_+}=0$ (or ${\partial t}/{\partial r_-}=0$),
which, by virtue of Eqs.~\eqref{HodTransEq}, implies
${\partial x}/{\partial r_+}=0$ (or ${\partial x}/{\partial r_-}=0$).
This means that this approach makes sense only in a region
where both Riemann invariants are position-dependent.

We look for the solution of the system (\ref{HodTransEq}) in the form
\begin{subequations}\label{HodTransEqSol}
\begin{align}
x-v_{+}(r_{+},r_{-})t & = w_+(r_+,r_-), \label{HodTransEqSol+} \\
x-v_{-}(r_{+},r_{-})t & = w_-(r_+,r_-). \label{HodTransEqSol-}
\end{align}
\end{subequations}
Inserting the above expressions in the system (\ref{HodTransEq})
yields, with account of Eqs.~(\ref{RiemannVel}),
\begin{equation}\label{eq-noname}
\begin{split}
\frac1{w_+-w_-}\frac{\prt w_+}{\prt r_-}=
\frac1{v_+-v_-}\frac{\prt v_+}{\prt r_-}, \\
\frac1{w_+-w_-}\frac{\prt w_-}{\prt r_+}=
\frac1{v_+-v_-}\frac{\prt v_-}{\prt r_+}.
\end{split}
\end{equation}
From expressions \eqref{RiemannVel}, \eqref{PhysVar_u} and
\eqref{PhysVar_rho} one sees that
${\partial v_+}/{\partial r_-} = {\partial v_-}/{\partial r_+}$. This
implies that $\prt w_+/\prt r_-=\prt w_-/\prt r_+$; we can thus
represent $w_{\pm}$ in term of a type of ``potential'' function
$W(r_+,r_-)$:
\begin{equation}\label{58-2}
  w_+=\frac{\prt W}{\prt r_+},\quad
  w_-=\frac{\prt W}{\prt r_-}.
\end{equation}
Inserting the expressions \eqref{58-2} into the system
\eqref{eq-noname} shows, with account of Eqs.~\eqref{RiemannVel},
\eqref{PhysVar_u} and \eqref{PhysVar_rho}, that $W$ is solution of the
Euler-Poisson equation
\begin{equation}\label{58-3}
\frac{\prt^2W}{\prt r_+\prt r_-}+
a(r_+,r_-)\frac{\prt W}{\prt r_+}+b(r_+,r_-)\frac{\prt W}{\prt r_-}=0,
\end{equation}
with
\begin{equation}\label{}
\begin{split}
& a(r_+,r_-)=-b(r_+,r_-) = \\
& =-\frac{1}{v_+-v_-}\frac{\partial v_+}{\partial r_-}=
 -\frac{1-c'(r_+-r_-)}{2c(r_+-r_-)},
\end{split}
\end{equation}
where $c'(r)=dc/dr$ and $c$ is computed as a function of $r=r_+-r_-$ from
expressions \eqref{LocalVel} and \eqref{PhysVar_rho}.

The characteristics of the second order partial differential equation
\eqref{58-3} are the straight lines $r_+=\xi=\mathrm{const}$ and
$r_-=\eta=\mathrm{const}$, parallel to the coordinates axis in the
hodograph plane. The Riemann method is based on the idea that one can
find the solution of the Euler-Poisson equation in a form similar to
d'Alembert solution of the wave equation, with explicit account of the
initial conditions which fix the value of $W$ on some curve
$\mathcal{C}$ in the hodograph plane. These initial data are
transferred along the characteristics into the domain of interest, so
that the function $W$ can be found at any point $P=(\xi,\eta)$.

Riemann showed (see, e.g., Refs.~\cite{Sommerfeld-49,CourantHilbert-62})
that $W(P)$ can be represented in the form
\begin{equation}\label{m1-282.9}
W(P)=\frac12(R{W})_A+\frac12(R{W})_B - \int_A^B(Vdr_++Udr_-),
\end{equation}
where the points $A$ and $B$ are projections of the
``observation'' point $P$
onto $\mathcal{C}$ along the $r_+$ and $r_-$ axis respectively. The
integral in \eqref{m1-282.9} is performed along $\mathcal{C}$, with
\begin{equation}\label{m1-284.4}
\begin{split}
& U=\frac12 \left(R\frac{\prt {W}}{\prt r_-}-
{W}\frac{\prt R}{\prt r_-}\right)+a{W}R,\\
& V=\frac12 \left({W}\frac{\prt R}{\prt r_+}
-R\frac{\prt {W}}{\prt r_+}\right)-b{W}R.
\end{split}
\end{equation}
$R(r_+,r_-;\xi,\eta)$ is the ``Riemann function'' which satisfies
the equation
\begin{equation}\label{eq16b}
\frac{\prt^2R}{\prt r_+\prt r_-}-a\frac{\prt R}{\prt r_+}-
b\frac{\prt R}{\prt r_-}-
\left( \frac{\partial a}{\partial r_+}+
\frac{\partial b}{\partial r_-}\right) R=0,
\end{equation}
with the additional conditions:
\begin{equation}\label{BoundR}
\begin{split}
\frac{\partial R}{\partial r_+} -bR = & 0
\quad \text{along the characteristic} \quad r_-=\eta, \\
\frac{\partial R}{\partial r_-} -aR = & 0
\quad \text{along the characteristic} \quad r_+=\xi, \\
\end{split}
\end{equation}
and $R(\xi,\eta;\xi,\eta)=1$.

One might think, looking at Eq.~(\ref{eq16b}), that we did not
progress towards the determination of the solution to
Eq.~(\ref{58-3}).  But we have replaced the initial
conditions for Eq.~(\ref{58-3}) by standard boundary conditions
(\ref{BoundR}) for Eq.~(\ref{eq16b}), independent of the initial
values of $\rho$ and $u$. The knowledge of the initial properties of
the flow is encapsulated in the known value of $W$ along ${\mathcal
  C}$, which appears in the right hand side of Eq. \eqref{m1-282.9}.

\subsection{General solution}\label{sec22}

We now consider a specific type of initial condition, for which
$u(x,0)=0$, with an initial parabolic bump profile
$\rho(x,0)=\orho(x)$ with
\begin{equation} \label{InitialParabola}
\begin{split}
\orho(x) =
\begin{cases}
\rho_0+(\rho_m-\rho_0)
\left(1-\frac{\displaystyle x^2}{\displaystyle l^2}\right),
& \; |x| \leq l, \\
\rho_0, & \; |x| > l.
\end{cases}
\end{split}
\end{equation}
Here $\rho_0$ is the background intensity, $l$ and $\rho_m$ are the
width and maximal intensity of the initial bump.
\begin{figure*}[ht]
  \centering
  \includegraphics[width=\linewidth]{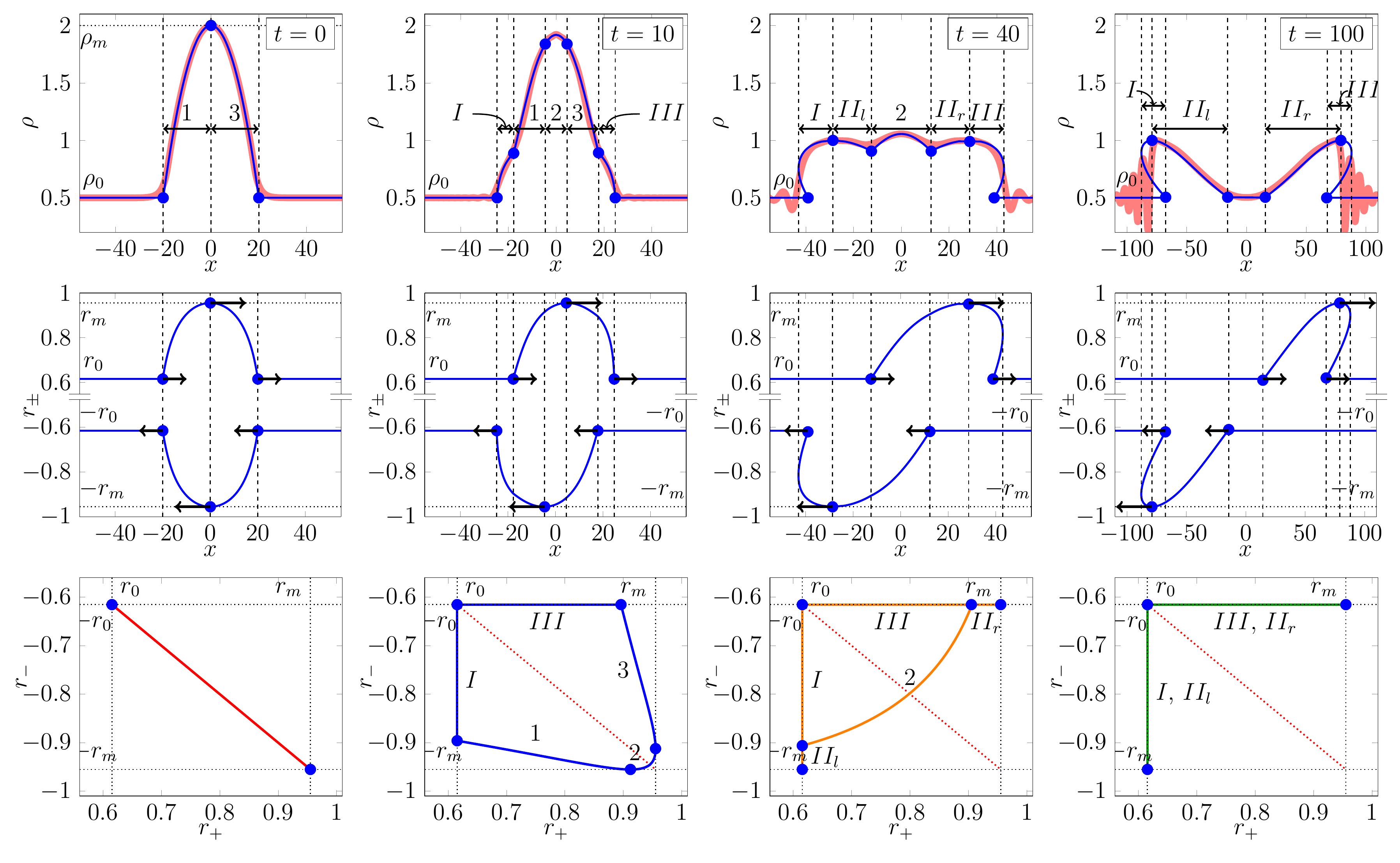}
  \caption{Behavior of characteristic quantities of the system.  Each
    column corresponds to a given value of $t$, each row to a
    different type of quantity. The top row displays the light
    intensity distribution $\rho(x,t)$ plotted as a function of
    $x$. Both the numerical solution of Eq.~(\ref{gNLS}) (red thick
    curves) and the analytic dispersionless solution (blue curves) are
    shown. The initial state is represented in the left panel and
    corresponds to Eq.~\eqref{InitialParabola} with $\rho_0=0.5$,
    $\rho_m=2$, $l=20$. The dynamics of the system is governed by
    Eq.~\eqref{gNLS} with $\gamma=1$. The middle row displays the
    (analytic result for) the Riemann invariants $r_+(x,t)$
    ($r_+\in[r_0,r_m]$) and $r_-(x,t)$ ($r_-\in[-r_m,-r_0]$) plotted
    as functions of $x$. The set of initial condition corresponds to
    $r_0=0.651$ and $r_m=0.955$.  Arrows indicate the direction of
    propagation. In the top and middle row the dashed lines divide
    space into several regions according to the behavior of the
    Riemann invariants. The Roman numbers indicate simple-wave
    regions, the Arabic ones correspond to regions in which both
    Riemann invariants are position-dependent.  The bottom row
    represents the behavior of Riemann invariants in the hodograph
    plane.  The dotted lines show the boundaries of the domain
    $[r_0,r_m]\times[-r_m,-r_0]$. Arabic and Roman numbers correspond
    to the notations of the regions from the top rows of the
    figure. Here the red curves correspond to $t=0$, the blue curves
    to $t=10$, the orange curves to $t=40$ and the green curves to
    $t=100$. The red dotted line represents the diagonal coinciding
    with the initial curve ${\cal C}$.}
  \label{fig:one}
\end{figure*}
This initial density profile is represented in the left panel on the
top row of Fig.~\ref{fig:one}, it is an even function of $x$;
generalization to non-symmetric distributions is straightforward. If
$l\gg\rho_m-\rho_0$, then deviations of the exact solution from its
hydrodynamic approximation are negligibly small almost everywhere,
except in small regions at the boundaries of the bump.

We denote the characteristic values of the Riemann invariant as $r_0$
and $r_m$ with
\begin{subequations}
  \begin{align}
  & r_0=r_+(|x|\ge l,t=0)=\frac{1}{\sqrt{\gamma}}
  \arctan{\sqrt{\gamma\rho_0}},\\
  & r_m=r_+(x=0,t=0)=\frac{1}{\sqrt{\gamma}}
  \arctan{\sqrt{\gamma\rho_m}},
 \end{align}
  \end{subequations}
see the left panel in the middle row of Fig.~\ref{fig:one}. It is
convenient to denote by $\overline{x}(r)$ the positive branch of the
reciprocal function of $r_+(x,t=0)$:
\begin{equation}\label{rdex}
r_+(x,0)=\tfrac12 \int_0^{\orho} \frac{c(\rho')}{\rho'}d\rho'.
\end{equation}
One gets
\begin{equation}\label{rhobder-parabol}
\orho(r_+)=\frac{1}{\gamma} \tan^2\left(\sqrt{\gamma} r_+\right),
\end{equation}
and
\begin{equation}\label{ox-parabol}
\overline{x}(r)=l
\sqrt{\frac{\rho_m-\orho(r)}{\rho_m-\rho_0}}.
\end{equation}
For the zero velocity initial profile we consider,
Eq.~\eqref{PhysVar_u} shows that the curve $\mathcal{C}$ which
represents the initial condition ($r_+(x,0),r_-(x,0)$) in the hodograph
plane is a segment of the antidiagonal $r_- = -r_+=-r$. Along it
Eqs.~(\ref{HodTransEqSol}) with $t = 0$ give
\begin{equation}\label{}
\left.\frac{\partial W}{\partial r_-}\right|_{\mathcal C}
=
\left.\frac{\partial W}{\partial r_+}\right|_{\mathcal C}=x.
\end{equation}
hence $W$ is constant along ${\cal C}$.
The value of this constant can be arbitrarily fixed to zero:
${W}(r,-r)=0$, and expressions (\ref{m1-284.4}) reduce to
\begin{equation}\label{}
U=\frac{x}{2}\, R(r,-r;\xi,\eta), \quad
V=-\frac{x}{2}\, R(r,-r;\xi,\eta).
\end{equation}

The top row of Fig.~\ref{fig:one} shows the light intensity
$\rho(x,t)$ plotted as a function of position for different $t$. The
middle row represents the corresponding distributions of the Riemann
invariants. At a given time, the $x$ axis can be considered as divided
into several domains, each requiring a specific treatment. Each domain
is characterized by the behavior of the Riemann invariants and is
identified in the upper row of Fig.~\ref{fig:one}. The domains in
which both Riemann invariants depend on $x$ are labeled by Arabic
numbers and the ones in which only one Riemann invariant depends on
$x$ are labeled by Roman numbers.  As was noted above, at the initial
moment $r_+=-r_-$, as can be seen, e.g., in the left panel of the
middle row of Fig.~\ref{fig:one}. Then, one of the Riemann invariants
begins to move in the positive direction of the $x$ axis, and the
other moves in the opposite direction. This behavior initially
leads to the configuration represented in the second column of
Fig.~\ref{fig:one}, where two simple-wave regions ($\rom{1}$ and
$\rom{3}$) and a new region (labeled ``2'') have appeared. At later
times (i.e., for longer sample lengths) region 2 persists while
regions $1$ and $3$ vanish and new simple-wave regions $\rom{2}_l$ and
$\rom{2}_r$ appear; this is illustrated in the third columns of
Fig.~\ref{fig:one}. At even larger lengths (right column of
Fig.~\ref{fig:one}), region $2$ also disappears and only simple-wave
regions persist: The initial pulse has completely split into two
pulses propagating in opposite directions.

It is worth noticing that the wave breaking corresponds to an overlap
between different regions which results in a multi-valued solution. If
we consider for instance the right propagating pulse, at the wave
breaking time $\twb$, region $\rom{3}$ starts overlapping with the (unnamed)
quiescent region where both Riemann invariants are constants, at the
right of the plot. Then these two regions both overlap with region
$\rom{2}_r$. These two configurations are respectively illustrated in
the columns $t=40$ and $t=100$ of Fig.~\ref{fig:one}. We will not dwell
on this aspect in detail, since a multi-valued solution is
nonphysical and, when dispersion is taken into account, it is replaced
by a dispersive shock wave (considered in Sec. \ref{sec4}).

\begin{figure}[ht]
\centering
\includegraphics[width=\linewidth]{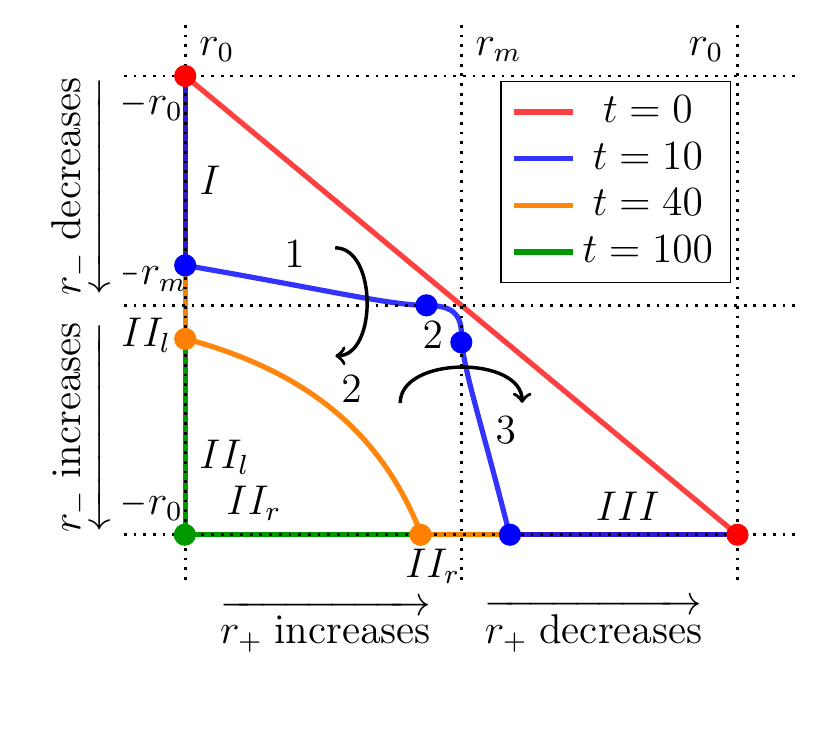}
\caption{Distributions of $r_+$ and $r_-$ in the four-sheeted
  hodograph plane (see the text).  The curves are the unfolded
  versions of the ones of the bottom row of Fig.~\ref{fig:one}.  The
  red line corresponds to the input distribution ($t=0$), while other
  colors correspond to other values of $t$, with the same color code
  as in the bottom row of Fig. \ref{fig:one}.  Curved arrows indicate
  the direction of unfolding of the domain
  $[r_0,r_m]\times[-r_m,-r_0]$.  The whole region above red line is
  unreachable for the initial distribution which we consider.}
\label{fig:three}
\end{figure}

The bottom row in Fig.~\ref{fig:one} represents the
behavior of $r_+$ and $r_-$ in the hodograph plane. Here, as before,
the simple-wave regions are indicated by Roman numerals while the
Arabic numerals correspond to regions where both Riemann invariants
depend on $x$. In each of the three domains 1, 2, and 3, the solution
$W$ of the Euler-Poisson equation has a different expression. In order
to describe these three branches, following Ludford \cite{Ludford-52},
we introduce several sheets in the characteristic plane by unfolding
the domain $[r_0,r_m]\times[-r_m,-r_0]$ into a four times larger
region as illustrated in Fig.~\ref{fig:three}. The potential
$W(r_+,r_-)$ can now take a different form in each of the regions
labeled $1$, $2$, and $3$ in Fig.~\ref{fig:three} and still be
considered as a single valued. From the relation (\ref{m1-282.9}) we can
obtain $W(\xi,\eta)$ in regions $1$ and $3$, by computing the right
hand side integral along the anti-diagonal ${\cal C}$,
between the points $A$ of
coordinates $(-\eta,\eta)$ and $B$ of coordinates $(\xi,-\xi)$:
\begin{equation}\label{W13}
\begin{split}
W^{(3)}(\xi,\eta)=-W^{(1)}(\xi,\eta)=\int_{-\eta}^{\xi}
\overline{x}(r)R(r,-r;\xi,\eta)dr
\end{split}
\end{equation}
The difference in signs in the above expressions of $W$ comes from the
fact that $x = \mp \overline{x}(r)$ depending on if one is in region
$1$ or $3$. For the region $2$, using unfolded surfaces (see
Fig.~\ref{fig:three} and also Ref. \cite{ikp-19c}) and upon
integrating by parts one obtains
\begin{equation}\label{W2}
\begin{split}
W^{(2)}(\xi,\eta)& =  \left(RW^{(1)}\right)_B+\left(RW^{(3)}\right)_A  \\
& +\int_A^C
\left.\left(\frac{\partial R}{\partial r_-}-aR\right)\right|_{r_+=r_m}
W^{(3)}dr_-  \\
& -\int_C^B
\left.\left(\frac{\partial R}{\partial r_+}-bR\right)\right|_{r_-=-r_m}
W^{(1)}dr_+,
\end{split}
\end{equation}
where the coordinates of the relevant points are:
$A=(r_m, \eta)$, $B=(\xi, -r_m)$ and $C=(r_m, -r_m)$.

From  the conditions (\ref{BoundR}) one can find that
\begin{equation}\label{}
\begin{split}
R(r_+,\eta;\xi,\eta) = & \sqrt{ \frac{c(\xi-\eta)}{c(r_+-\eta)} }
\exp{\left(\frac{1}{2}
\int_{\xi-\eta}^{r_+-\eta} \!\!\!\! \frac{dr}{c(r)}\right)}, \\
R(\xi,r_-;\xi,\eta) = & \sqrt{ \frac{c(\xi-\eta)}{c(\xi-r_-)} }
\exp{\left(\frac{1}{2}
\int_{\xi-\eta}^{\xi-r_-} \!\!\!\! \frac{dr}{c(r)}\right)}.
\end{split}
\end{equation}
These expressions suggest that $R$ can be looked for in the form
\begin{equation}\label{RFull}
\begin{split}
R(r_+,r_-;\xi,\eta) = & \\
\sqrt{ \frac{c(\xi-\eta)}{c(r_+-r_-)} } &
\exp{\left(\frac{1}{2} \int_{\xi-\eta}^{r_+-r_-} \!\!\!\!
\frac{dr}{c(r)}\right)} F(r_+,r_-;\xi,\eta),
\end{split}
\end{equation}
where $F(r_+,\eta;\xi,\eta)=F(\xi,r_-;\xi,\eta)=1$.  For small enough
$t$, when $\xi$ is close to $r_m$ and $\eta$ is close to $-r_m$, the
integrand functions in Eq.~(\ref{W2}) are small by virtue of
Eqs.~(\ref{BoundR}); accordingly $F \simeq 1$ in
Eq.~(\ref{RFull}).  Such an approximation has been already used
in Ref.~\cite{ikp-19a} for
discarding the integrated terms in the right-hand side of
Eq.~(\ref{W2}), and showed good applicability to the
situation when $\gamma=0$.
Thus, using this approximation, we obtain an approximate
expression for the Riemann function (see also Ref. \cite{ikp-19c}):
\begin{equation}\label{}
R(r_+,r_-;\xi,\eta)  \simeq \mathscr{R}(r_+-r_-,  \xi-\eta),
\end{equation}
where
\begin{equation}\label{curlyR}
\begin{split}
\mathscr{R}(r_1,r_2)&=\sqrt{ \frac{c(r_2)}{c(r_1)} }
\exp{\left(\frac{1}{2} \int_{r_2}^{r_1} \!\! \frac{dr}{c(r)}\right)}
\\
&=\sqrt{\frac{c(r_2) \rho(r_1)}{c(r_1) \rho(r_2)}},
\end{split}
\end{equation}
where $\rho$ is expressed as a function of $r_1$ (or $r_2$) through Eq.
\eqref{PhysVar_rho}. A simple calculation yields
\begin{equation}\label{curlyR_gamma}
\mathscr{R}(r_1,r_2)=
\sqrt{\frac{
\sin\left(\frac{\sqrt{\gamma}}{2}r_1\right)
\cos^3\left(\frac{\sqrt{\gamma}}{2} r_2\right)}
{\sin\left(\frac{\sqrt{\gamma}}{2}r_2\right)
\cos^3\left(\frac{\sqrt{\gamma}}{2}r_1\right)}}.
\end{equation}
We can thus adapt expressions \eqref{W13} for $W$ in regions $1$ and $3$
\begin{equation}\label{W13fin}
\begin{split}
W^{(1)}&(r_+,r_-) =  -W^{(3)}(r_+,r_-) =\\
& - \int_{-r_-}^{r_+}\!\!\! \overline{x}(r)\mathscr{R}(2r,r_+-r_-)dr, \\
\end{split}
\end{equation}
whereas expression \eqref{W2} in region $2$ now reads
\begin{equation}\label{W2fin}
\begin{split}
W^{(2)}(r_+,r_-)= &  \mathscr{R}(r_m-r_-,r_+-r_-)
W^{(3)}(r_m,r_-) \\
+& \mathscr{R}(r_++r_m,r_+-r_-)
W^{(1)}(r_+,-r_m).
\end{split}
\end{equation}
In expressions \eqref{W13fin} and \eqref{W2fin} we have made, with
respect to Eqs.~\eqref{W13} and \eqref{W2}, the replacements
$\xi \rightarrow r_+$, $\eta \rightarrow r_-$ so that these
expressions can be used in Eqs.~(\ref{58-2}) and then
(\ref{HodTransEqSol}). The knowledge of the expression of $W$ is the
regions where both Riemann invariants vary (regions 1,2 and 3) makes
it possible to compute $r_{\pm}(x,t)$ in these regions, as detailed in
Ref. \cite{ikp-19c}.  The density and velocity profiles are then
obtained from Eqs.~(\ref{PhysVar_u}) and \eqref{PhysVar_rho}.

\subsection{Simple wave solution}\label{sec23}

In the simple-wave regions, where one of the Riemann invariants is
constant, the hodograph transform does not apply. In such a region
however, the solution of the problem is relatively easy. For instance,
during the initial stages of evolution, when the regions $\rom{1}$ and
$\rom{3}$ still exist, we can solve the problem by means of the method
of characteristics. We look for a solution in the form
\begin{equation}\label{}
\begin{split}
x - v_+(r_+,-r_0)t & = h^{\rom{3}}(r_+), \qquad \text{for region $\rom{3}$},\\
x - v_-(r_0,r_-)t & = h^{\rom{1}}(r_-), \qquad \text{for region $\rom{1}$}.
\end{split}
\end{equation}
where the functions $h^{\rom{3}}$ and $h^{\rom{1}}$ are determined by
boundary conditions: the simple wave solution should match with the
solution in regions where both Riemann invariants vary at the
interface between the two regions ($\rom{3}$ and 3 or $\rom{1}$ and
1). This yields from Eqs.~(\ref{HodTransEqSol})
\begin{equation}\label{swrIII}
\begin{split}
x - v_+(r_+,-r_0)t = \frac{\partial W^{(3)}(r_+,-r_0)}{\partial r_+},
\end{split}
\end{equation}
for the simple-wave region $\rom{3}$, and
\begin{equation}\label{}
\begin{split}
x - v_-(r_0,r_-)t = \frac{\partial W^{(1)}(r_0,r_-)}{\partial r_-},
\end{split}
\end{equation}
for region $\rom{1}$.

After a certain time, two new simple-wave regions appear, which are
denoted as $\rom{2}_l$ and $\rom{2}_r$ in Fig.~\ref{fig:one}).
Similarly, we get
\begin{equation}\label{swrIIr}
\begin{split}
x - v_+(r_+,-r_0)t = \frac{\partial W^{(2)}(r_+,-r_0)}{\partial r_+},
\end{split}
\end{equation}
for region $\rom{2}_r$, and for region $\rom{2}_l$:
\begin{equation}\label{swrIIl}
\begin{split}
x - v_-(r_0,r_-)t = \frac{\partial W^{(2)}(r_0,r_-)}{\partial r_-}.
\end{split}
\end{equation}

From the results presented in Secs.~\ref{sec22} and \ref{sec23} we
obtain a complete description of the dispersionless stage of evolution
of the system. The top row of Fig.~\ref{fig:one} displays a comparison
of the results of this approach with the numerical solution of
Eq.~(\ref{gNLS}). There is a very good agreement up to the wave
breaking moment. For subsequent time ($t>\twb$) the dispersionless evolution
becomes multi-valued in some regions of space (indicating a breakdown
of the approach and the occurrence of a DSW), but remains accurate in
others which we denote as ``dispersionless regions'' below. Note that
these regions can be
precisely defined only after the extension of the DSW has been
properly determined. This question is addressed in the following section.

\section{Dispersive shock waves}\label{sec4}

The basic mathematical approach for the description of DSWs is based
on the Whitham modulation theory \cite{Whitham-65,Whitham-74}.  This
treatment relies on the large difference between the fast oscillations
within the wave and the slow evolution of its envelope. It results in
the so-called Whitham modulation equations which constitute a very
complex system of nonlinear first order differential
equations. Whitham's great achievement was to be able to transform
such a system --- in the very important and universal case of the
Korteweg-de Vries (KdV) equation --- into a diagonal (Riemann) form
analogous to the system \eqref{RiemannEquat}. This achievement enabled
Gurevich and Pitaevskii \cite{gp-73} to successfully apply the Whitham
theory to the description of DSWs dynamics for the KdV equation in a
system experiencing simple wave breaking. It became clear later that
the possibility to diagonalize the system of Whitham equations is
closely related with the complete integrability of the KdV
equation, and Whitham theory was extended to many completely
integrable equations, as described, for example, in the reviews
\cite{DubrovinNovikov-93} and \cite{Kamchatnov-97}.

However, a large number of equations are not completely integrable
--- among which Eq.~\eqref{gNLS} --- and require the development of a more
general theory for describing DSWs.  An important success in this
direction was obtained by El who developed in Ref. \cite{El-05} a
method that made it possible to find the main parameters of a DSW
arising from the evolution of an initial discontinuity. Recently, it
was shown in Ref.~\cite{Kamchatnov-19} that El's method can be
generalized to a substantial class of initial conditions, such as
simple waves, for which the limiting expressions for the
characteristic velocities of the Whitham system at the edges of the
DSW are known from general considerations as being equal to either the
group velocity of the wave at the boundary with a smooth solution, or
the soliton velocity at the same boundary. This information, together
with the knowledge of the smooth solution in the
dispersionless regions, is enough to find the law of motion of the
corresponding edge of the DSW for an arbitrary initial profile. We
will use the methods of Ref.~\cite{Kamchatnov-19} to find the dynamics
of the edges of the DSW in our case.

The specific dispersive properties of the system under consideration enter
into the general theory in the form of Whitham's ``number of waves''
conservation law \cite{Whitham-65,Whitham-74}
\begin{equation}\label{eq1}
\frac{\prt k}{\prt t}+\frac{\prt \om(k)}{\prt x}=0,
\end{equation}
where $k=2\pi/L$ and $\om=kV$ are the wave vector and the angular
frequency of a single phase nonlinear wave, $L$ being its wavelength
and $V$ its phase velocity. Because of the large difference between
the scales characterizing the envelope and those characterizing the
oscillations within the wave, Eq.~\eqref{eq1} is valid along the whole
DSW.  At the small amplitude edge of the DSW the function $\om(k)$
becomes the linear dispersion relation $\om_{\rm lin}(k)$ in the
system. In our case, this dispersion law is obtained by linearizing
equations (\ref{Hydro}) around the uniform state $\rho=\rho_0$,
$u=u_0$ (we keep here a nonzero value of $u_0$ for future
convenience); that is, we write $\rho(x,t)=\rho_0+\rho'(x,t)$ and
$u(x,t)=u_0+u'(x,t)$, where $|\rho'|\ll\rho_0$ and $|u'|\ll u_0$. A
plane wave expansion of $\rho'$ and $u'$ immediately yields
\begin{equation}\label{linlaw}
\omega_{\rm lin}(k) \equiv
k u_0 \pm k \sqrt{\frac{\rho_0}{(1+\gamma\rho_0)^2}+\frac{k^2}{4}}.
\end{equation}
Thus, if one can calculate the wave number at the small-amplitude edge of
the DSW, one obtains the speed of propagation of this edge
as being equal to the corresponding group
velocity $v_g(k)=d\omega_{\rm lin}/dk$.

Unfortunately, this approach cannot be applied straightforwardly at
the soliton edge of a DSW. In spite of that, El showed \cite{El-05}
that under some additional assumptions one can obtain from
Eq.~(\ref{eq1}) an ordinary differential equation relating the
physical variables along the characteristic of Whitham equations at
the soliton edge of the DSW. The idea is based on
the following remark: At the tails of a soliton, the density profile
has an exponential form $\propto \exp(-\tk|x - V_s t|)$, where $V_s$
is the speed of the soliton. Of course a soliton propagates at the
same velocity as its tails, which, being a small perturbations of the
background, obey the linear dispersion law $\om = \om_{\rm lin}(k)$.
Hence, as was noticed by Stokes~\cite{stokes}
(see also Sec.~252 in \cite{lamb}),
we arrive at the statement that the soliton velocity $V_s$ is
related to its ``inverse half-width'' $\tk$ by the formula
\begin{equation}\label{eq2}
V_s={\tom(\tk)}/{\tk},\quad\mbox{where}\quad
\tom(\tk)\equiv-i\om_{\rm lin}(i\tk).
\end{equation}
However, the ``solitonic counterpart''
\begin{equation}\label{eq4}
\frac{\prt \tk}{\prt t}+\frac{\prt \tom(\tk)}{\prt x}=0
\end{equation}
of Eq.~(\ref{eq1}) does not apply in all possible DSW configurations,
even at the soliton edge. Interestingly, it does apply for a step-like
type of initial conditions, and El used this property to determine the
inverse half-width and velocity of a soliton edge, using a procedure
similar to the one employed at the small amplitude edge. As a result,
a number of interesting problems were successfully considered by this
method, see, e.g.,
Refs.~\cite{egkkk-07,ElGrimshaw-06,ElGrimshaw-09,EslerPearce-11,Hoefer-14,CKP-16,HEK-17,AnMarchant-18}.

To go beyond the initial step-like type of problems, one has to use
some additional information about the properties of the Whitham
modulation equations at the edges of DSWs (see
Refs. \cite{Kamchatnov-19,IvanovKamchatnov-2019}).  For integrable
equations, it is known that the ``soliton number of waves conservation
law'' (\ref{eq4}) is valid in the case where the DSW is triggered by a
simple wave breaking \cite{Kamchatnov-19}. Therefore, it is natural to
assume that Eq.~\eqref{eq4}
also applies for non-integrable equations in situations
where the pulse considered propagates into a uniform and stationary
medium\footnote{More precisely: propagates into a medium for which
  $r_+$ and $r_-$ are constant.} and experiences a simple wave
breaking.

We can consider that the smooth solution of the dispersionless
equations is known from the approach presented in Sec.~\ref{sec3}.  At
the boundary between the dispersionless simple wave region and the
DSW, Eqs.~(\ref{eq1}) and (\ref{eq4}) reduce to ordinary differential
equations which can be extrapolated to the whole DSW.  Their solution,
with known boundary condition at both edges, yields the wave number
$k$ and the inverse half-width of solitons $\tk$ at the boundary with
the smooth part of the pulse. Consequently, the corresponding group
velocity or the soliton velocity at a DSW edge can be expressed in
terms of the parameters of the smooth solution at its boundary with a
DSW.  These velocities can be considered as the characteristic
velocities of the limiting Whitham equations at this edge, which makes
it possible to represent these equations in the form of first order
partial differential equations after a hodograph transformation.  The
compatibility condition for this partial differential equation with a
smooth dispersionless solution gives the law of motion of this DSW
edge.  If the soliton solution of the nonlinear equation at hand is
known, then the knowledge of its velocity makes it possible to also
determine its amplitude at the boundary of the DSW. In this paper, we
shall apply this method to study the evolution of initial simple-wave
pulses in the generalized NLS equation \eqref{gNLS}.

We suppose that in the whole region of the DSW, one has
\begin{equation} \label{eq017}
\begin{split}
r_-& =\frac{u}{2}-\frac{1}{\sqrt{\gamma}}\arctan{\sqrt{\gamma\rho}}\\
& =
-\frac{1}{\sqrt{\gamma}}\arctan{\sqrt{\gamma\rho_0}}=-r_0=\rm{cst}.
\end{split}
\end{equation}
This occurs for instance in the DSW formed at the right of the right
propagating bump issued from the initial condition
\eqref{InitialParabola} --- for which the behavior of the
dispersionless Riemann invariants is sketched in Fig.~\ref{fig:one}
--- and also in the model cases studied in Secs.~\ref{sec411} and
\ref{sec42} below.

Equation (\ref{RiemannEquat-}) is satisfied identically by virtue of
our assumption \eqref{eq017}. Also, since $r_-$ is constant, $r_+$ and
$u$ can be considered as functions of $\rho$ only [see
Eqs. \eqref{PhysVar_u} and \eqref{crprm}]:
\begin{equation} \label{eq17}
\begin{split}
u(\rho)&=\frac{2}{\sqrt{\gamma}}\left(\arctan{\sqrt{\gamma\rho}}
-\arctan{\sqrt{\gamma\rho_0}}\right), \\
r_+(\rho)& =\frac{1}{\sqrt{\gamma}}\left(2\arctan{\sqrt{\gamma\rho}}
-\arctan{\sqrt{\gamma\rho_0}}\right),
\end{split}
\end{equation}
and Eq.~\eqref{RiemannEquat+} for $r_+$ thus takes the form
\begin{equation} \label{eq18}
\frac{\prt \rho}{\prt t}+(u(\rho)+c(\rho))\frac{\prt \rho}{\prt x}=0,
\end{equation}
where the expression of the
local sound velocity is given in (\ref{LocalVel}).
The general solution \eqref{HodTransEqSol+} of Eq.~\eqref{RiemannEquat+}
here specializes to
\begin{equation} \label{eq19}
x-(u(\rho)+c(\rho))t =w_+(r_+(\rho),-r_0).
\end{equation}
All the relevant characteristic quantities of the DSW formed after the
wave breaking of such a pulse can be considered to be functions of
$\rho$ only.
We can thus write the right propagating
soliton dispersion law \eqref{eq2} in the form
\begin{equation}\label{omegatild}
  \begin{split}
\tom(\tk) & = \tk \left[u(\rho) +\sqrt{c^2(\rho)-\tfrac14 \tk^2}\right]\\
& =\tk \left[ u(\rho) + c(\rho) \tal(\rho) \right],
\end{split}
\end{equation}
where
\begin{equation}\label{}
\tal(\rho) = \sqrt{1-\frac{\tk^2}{4\, c^2(\rho)}}.
\end{equation}
We then have
\begin{equation}\label{ktild}
\tk(\rho)=2\,c(\rho)\, \sqrt{1-\tal^2(\rho)},
\end{equation}
where $\tal(\rho)$ is yet an unknown function.

As implied by El's method, along the soliton edge of the
 DSW one considers that $r_-$ is also a constant, and
  that, at all time, the quantities $u$, $c$, $r_+$, $\tk$, $\tal$ and
  $\tom$ can be considered as functions of $\rho$ only.
Then the solitonic ``number of wave
conservation'' (\ref{eq4}) yields, with account of (\ref{eq18}):
\begin{equation}
\frac{d \,\tom}{d\rho}=[u(\rho)+c(\rho)]\frac{d \,\tk}{d\rho}.
\end{equation}
Substitution of \eqref{omegatild} and (\ref{ktild}) into this equation
yields an ordinary differential equation for $\tal(\rho)$,
\begin{equation}\label{9.4}
\frac{d\tal}{d\rho}=-
\frac{(1+\tal)(1+3\gamma\rho+2\tal(1-\gamma\rho))}
{2\rho(1+\gamma\rho)(1+2\tal)},
\end{equation}
which should be solved with the boundary condition appropriate to the
situation considered (see below).

The same method is employed for studying the motion of the small
amplitude edge of the DSW. This edge propagates at the linear group
velocity, and for determining the relevant wave-vector, we rewrite
equation (\ref{linlaw}) in the form
\begin{equation}\label{omega}
\om_{\rm lin}(k) = k \left[ u(\rho) + c(\rho) \al(\rho) \right],
\end{equation}
that is
\begin{equation}\label{}
\al(\rho) = \sqrt{1+\frac{k^2}{4\,c^2(\rho)}},
\end{equation}
and
\begin{equation}\label{k}
k(\rho)=2\,c(\rho)\sqrt{\al^2(\rho)-1}.
\end{equation}
A simple calculation leads to the following differential equation
\begin{equation}\label{9.5}
\frac{d\al}{d\rho}=-
\frac{(1+\al)(1+3\gamma\rho+2\al(1-\gamma\rho))}{2\rho(1+\gamma\rho)(1+2\al)},
\end{equation}
which actually coincides with the equation (\ref{9.4}).  Now we
``extrapolate'' this equation across the entire DSW an determine the
wave-vector as the value of $k(\rho)$ at the density of the small
amplitude edge. The appropriate boundary condition for solving
\eqref{9.5} is fixed at the solitonic edge, and depends on the
configuration under study (see below).

\subsection{Positive pulse}\label{sec41}

In all this section we consider an initial condition with a region of
increased light intensity over a stationary uniform background.

\subsubsection{Constant Riemann invariant $r_-$}\label{sec411}

\begin{figure}[t!]
	\centering
	\includegraphics[width=\linewidth]{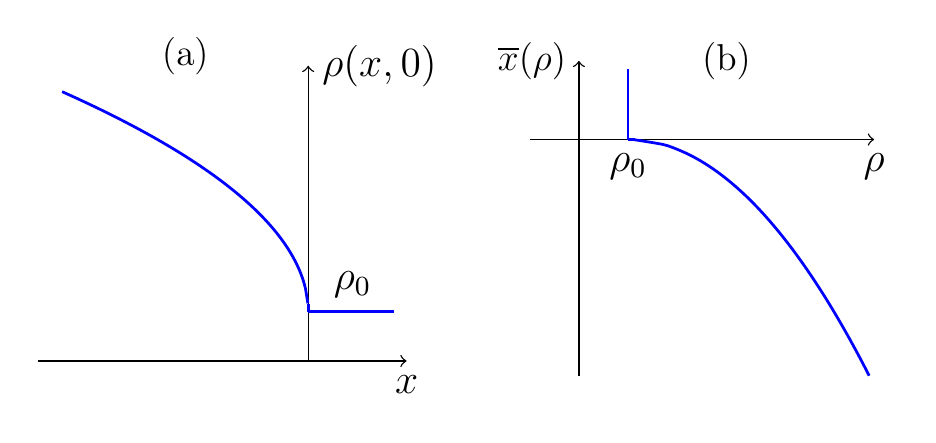}
	\caption{(a) Initial profile $\orho(x)$ of a monotonous
          positive pulse. (b) Inverse function $\overline{x}(\rho)$.}
	\label{fig:four}
\end{figure}

As a first illustration of the method we consider an initial
condition for which the Riemann invariant $r_-$ has the constant value
$-r_0$ for all $x$.  We shall start with a pulse with the following initial
intensity distribution:
\begin{equation}\label{c_init}
\orho(x)=
\begin{cases}
\rho_0+\widetilde{\rho}(x) & \mbox{if}\quad x<0, \\
\rho_0 & \mbox{if}\quad x\ge 0,
\end{cases}
\end{equation}
where $\widetilde{\rho}(x)$ is a decreasing function of $x$ with a
vertical tangent at $x\to 0^-$, see Fig.~\ref{fig:four}(a).
From \eqref{RiemannInv} one then obtains
$u(x,0)=\ovu(x)$ with
\begin{equation}\label{u_init}
  \ovu(x)=
  \frac{2}{\sqrt{\gamma}}\arctan\sqrt{\gamma \orho(x)} - 2\, r_0\; .
  \end{equation}
This non-standard type of initial condition is of interest because it
enables to test the theory just presented in a particularly simple
setting in which (i) the evolution of the non-dispersive part of the
system is that of a simple wave, i.e., one can replace $w_+$ in the
right hand side of Eq.~\eqref{eq19} by $\ox(\rho)$, where $\ox(\rho)$
is the function inverse to the initial distribution of the intensity
$\orho(x)$ (see Fig.~\ref{fig:four}(b)), (ii) the wave breaking occurs
instantaneously, at the front edge of the pulse. This wave breaking is
followed by the formation of a DSW with a small-amplitude wave at its
front edge and with solitons at its trailing edge, at the boundary
with the smooth part of the pulse [described by the equation
  (\ref{eq19})].  Therefore, we can find the law of motion of the rear
solitons edge of the DSW by the method of Ref.~\cite{Kamchatnov-19}.

Eq.~\eqref{9.4} should be solved with the boundary condition
$\tal(\rho_0) = 1$, that is, the soliton inverse half-width vanishes
together with its amplitude at the small-amplitude edge.

When the function $\tal(\rho)$ is known, the
velocity of the front edge can be represented as the soliton velocity
\begin{equation}\label{vsderhos}
V_s=\frac{\tom}{\tk}=u(\rho_s)+c(\rho_s)\tal(\rho_s) ,
\end{equation}
where $\rho_s$ is the density at the solitonic edge of the DSW.  We
notice again that $V_s$ is the characteristic velocity of the Whitham
modulation equations at the soliton edge. The system being
nonintegrable, the corresponding Whitham equation is not known, but its
limiting form at the solitonic edge can be written explicitly: along
this boundary one has $dx_s-V_s dt=0$, where $x_s(t)$ is the position
of the solitonic edge at time $t$ and $\rho_s=\rho(x_s(t),t)$.  It is
convenient to re-parametrize all the quantities in term of $\rho_s$:
$x(\rho_s)$ and $t(\rho_s)$. This leads to
\begin{equation} \label{}
\frac{\prt x_s}{\prt \rho_s}-V_s(\rho_s) \frac{\prt t}{\prt \rho_s}=0.
\end{equation}
This equation should be compatible with Eq.~(\ref{eq18}) at the
solitonic boundary between the DSW and the dispersionless region
gives, with account of (\ref{eq19}). Here, the specific initial
condition we have chosen (with $r_-(x,0)=\rm{cst}$)
simplifies the problem because one has
$w_+(r_+(\rho),-r_0)=\bar{x}(\rho)$. The compatibility equation
then reads
\begin{equation} \label{xxx}
\frac{\sqrt{\rho_s}}{1+\gamma\rho_s}(1-\tal)\frac{d t}{d \rho_s}+
\frac{3+\gamma\rho_s}{2\sqrt{\rho_s}(1+\gamma\rho_s)^2}t=
-\frac{d\ox(\rho_s)}{d \rho_s}.
\end{equation}
One sees from Fig.~\eqref{fig:four} that for the initial condition we
consider, the wave breaking occurs instantly at $t=0$, with
$\rho_s=\rho_0$. Hence Eq.~\eqref{xxx} should be integrated with the
initial condition $t(\rho_0) = 0$. The corresponding solution reads
\begin{equation} \label{tParam}
t(\rho_s)=\int_{\rho_0}^{\rho_s}
\frac{(1+\gamma\rho)\ox'(\rho)G(\rho_s,\rho)}{\sqrt{\rho}(\tal(\rho)-1)}
d\rho,
\end{equation}
where $\ox'=d\ox/d\rho$ and
\begin{equation} \label{GParam}
G(\rho_s,\rho)=\exp{\left(\int_{\rho}^{\rho_s}
\frac{3+\gamma\rho'}{2\rho'(1+\gamma\rho')(\tal(\rho')-1)}d\rho'\right)}.
\end{equation}
Consequently,
\begin{equation}\label{xParam}
x_s(\rho_s) = \left[u(\rho_s)+c(\rho_s)\right]t(\rho_s) +\ox(\rho_s),
\end{equation}
where $u(\rho)$ and $c(\rho)$ are given by Eqs.~\eqref{eq17} and
\eqref{LocalVel}, respectively.

\begin{figure}[t]
	\centering
	\includegraphics[width=0.5\textwidth]{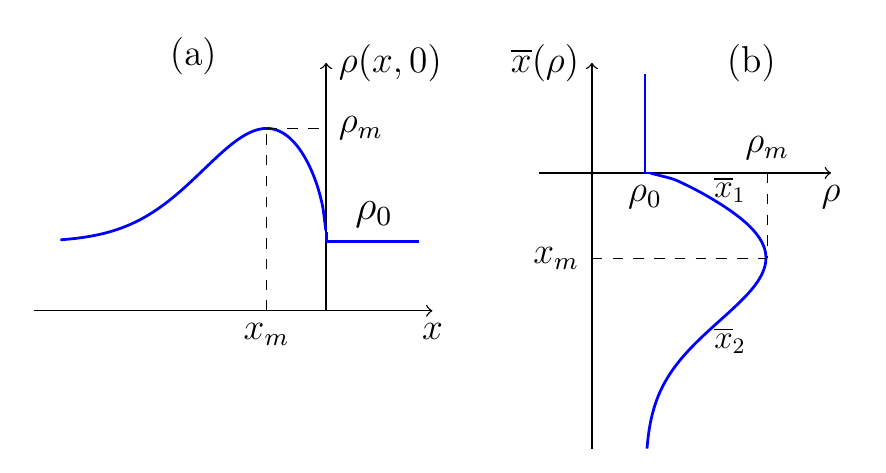}
	\caption{(a) Initial profile $\orho(x)$ of a non-monotonous
          positive pulse.  (b) The inverse function
          $\overline{x}(\rho)$ is represented by two branches
          $\overline{x}_1(\rho)$ and $\overline{x}_2(\rho)$. Here
          $x_m$ is the point where $\rho$ reaches its maximal value
          $\rho_m$.}
	\label{fig:five}
\end{figure}

As a generalisation of the model initial profile represented in
Fig. \ref{fig:four}, we now consider the case where the initial
distribution $\overline{\rho}(x)$ is non monotonous, with a single
maximum. This amounts to assume that $\widetilde{\rho}$ in
\eqref{c_init} has a maximum $\rho_m$ at $x=x_m<0$, tends to 0 fast
enough as $x\to-\infty$, with still a vertical tangent at $x\to 0^-$,
see Fig.~\ref{fig:five}(a). This last condition means that we suppose
again for convenience that the wave breaks at the moment $t=0$. The
reciprocal function of $\orho(x)$ becomes now two-valued and we denote
its two branches as $\ox_1(\rho)$ and $\ox_2(\rho)$, see
Fig.~\ref{fig:five}(b). In an initial stage of development of the DSW,
the formulas obtained previously for a monotonous initial condition
straightforwardly apply. This occurs when the maximum of the smooth
part of the profile issued from the initial distribution has not yet
penetrated the DSW. In this case the soliton edge propagates along the
branch $\ox_1(\rho)$ and one should just use this function instead of
$\ox(\rho)$ in Eqs. \eqref{tParam} and \eqref{xParam}.  A second stage
begins when the maximum of the smooth part of the profile reaches the
DSW. In this case one can show\footnote{See the detailed study of a
  similar situation in Sec. \ref{sec412}.} that
Eqs. \eqref{tParam} and \eqref{xParam} are modified to
\begin{subequations} \label{tParamLP}
\begin{align}
t(\rho_s) & =\int_{\rho_0}^{\rho_m}
\frac{(1+\gamma\rho)\ox_1'(\rho)G(\rho_s,\rho)}{\sqrt{\rho}(\tal(\rho)-1)}
d\rho \nonumber \\
 & \quad + \int_{\rho_m}^{\rho_s}
 \frac{(1+\gamma\rho)\ox_2'(\rho)G(\rho_s,\rho)}{\sqrt{\rho}(\tal(\rho)-1)}
 d\rho, \label{tParamLPa} \\
x_s(\rho_s) &= \left[u(\rho_s)+c(\rho_s)\right]t(\rho_s) +\ox_2(\rho_s).
\label{tParamLPb}
\end{align}
\end{subequations}

To determine the position of the small-amplitude edge we solve
Eq.~(\ref{9.5}) with the boundary condition
\begin{equation}\label{7.7}
\al(\rho_m)=1,
\end{equation}
which corresponds to the moment at which the soliton edge reaches the
maximal point of the initial distribution.  The corresponding solution
$\al(\rho)$ yields the spectrum \eqref{k} of all possible wave numbers
$k(\rho)$, with $\rho$ ranging from $\rho_0$ to $\rho_m$.  The maximal
value of the group velocity $d\om_{\rm lin}/dk$ is reached at
$\rho=\rho_0$ and it provides the asymptotic value of the velocity of
the small amplitude edge:
\begin{equation}\label{v_small_amp}
\frac{dx_r}{dt}=\left.\frac{d\omega_{\rm lin}}{dk}\right|_{\rho=\rho_0}=
\frac{\sqrt{\rho_0}}{1+\gamma\rho_0}\frac{2\al^2(\rho_0)-1}{\al(\rho_0)}.
\end{equation}

\begin{figure}[t]
	\centering
	\includegraphics[width=\linewidth]{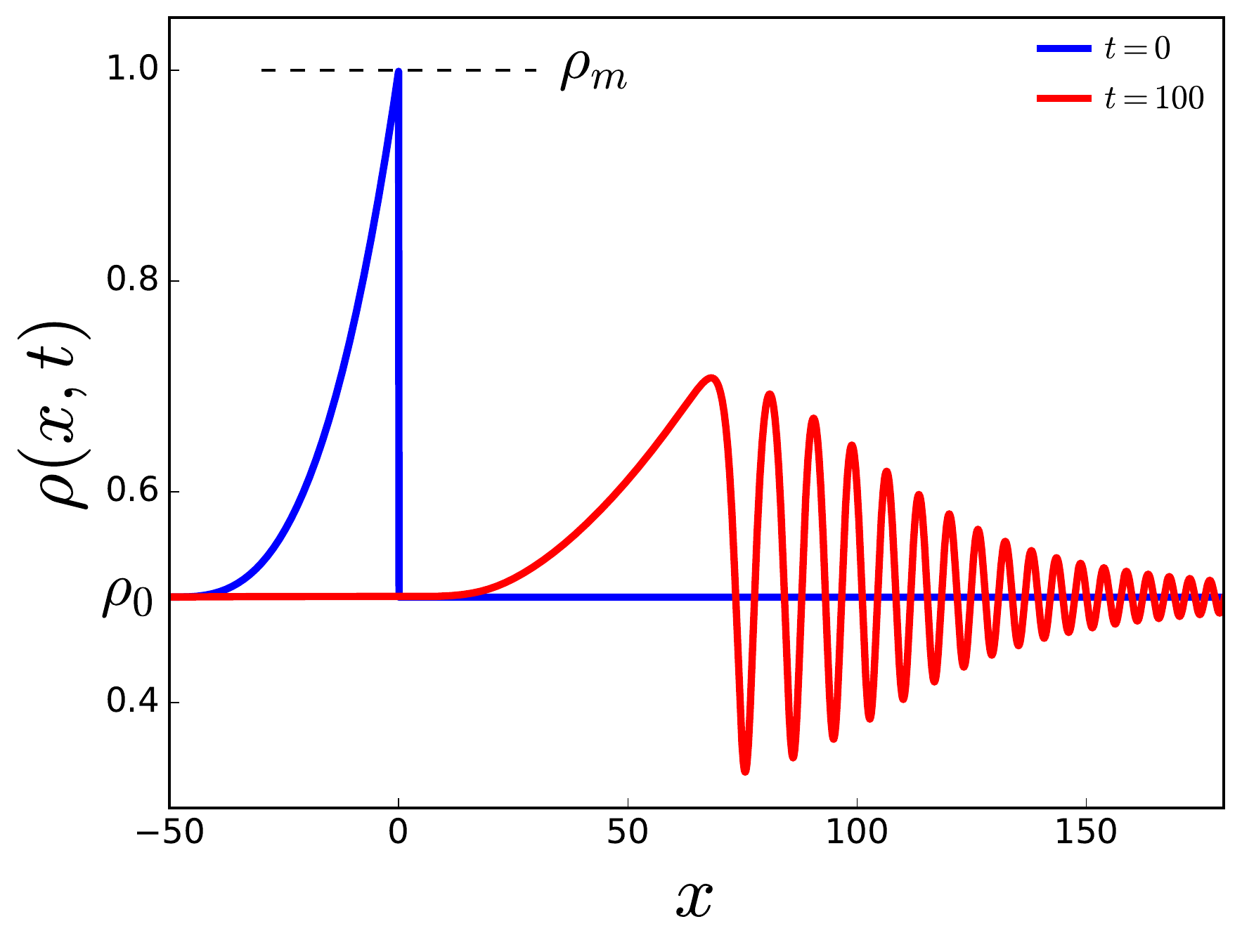}
	\caption{Blue solid curve: simple-wave initial condition
          (\ref{InitialFoeDSWPos}) with $\rho_0=0.5$, $\rho_m=1$ and
          $x_0=50$. Red solid curve: resulting dispersive shock
          wave computed by numerically solving Eq. \eqref{gNLS}
           with $\gamma=1$.}
	\label{fig:six}
\end{figure}

As an illustration of the accuracy of the method, we now compare the
theoretical results with those of a numerical solution of the
generalized NLS equation (\ref{gNLS}) for the non-monotonous initial
intensity distribution
\begin{equation}\label{InitialFoeDSWPos}
\orho(x) =
\begin{cases}
\rho_0+(\rho_m-\rho_0) \left(\frac{x}{x_0}+1\right)^3
\; \mbox{if}\; x\in[-x_0,0],\\
\rho_0 \quad\mbox{elsewhere},
\end{cases}
\end{equation}
see the blue curve in Fig.~\ref{fig:six}. The typical form of the DSW
generated by such a pulse is represented in the same figure by a red
curve.  For the particular initial profile \eqref{InitialFoeDSWPos}
the branch $\overline{x}_1(\rho)$ shrinks to zero, the first term in
the right hand side of \eqref{tParamLPa} cancels and we need take into
account only the contribution of the second branch, given by the
expression
\begin{equation}\label{}
\overline{x}_2(\rho\ge\rho_0) = x_0
\left( \frac{\rho-\rho_0}{\rho_m-\rho_0} \right)^{1/3} - x_0.
\end{equation}
For the numerical simulation, we took instead of the idealized profile
\eqref{InitialFoeDSWPos} the numerical initial condition:
\begin{equation}\label{gen20}
  \overline\rho(x)=\rho_0+(\rho_m-\rho_0)\left(1+x/x_0\right)^3
  \frac{1-\tanh(x/w)}{2}\; ,
\end{equation}
with $w=1$, $x_0=50$, $\rho_0=0.5$ and $\rho_m=1$. Equation
\eqref{gen20} yields a maximum value of $\overline\rho$ which is not
exactly equal to $\rho_m=1$, but to $\overline\rho_{\rm max}=0.9362$.
\begin{figure}[t!]
  \centering
  \includegraphics[width=\linewidth]{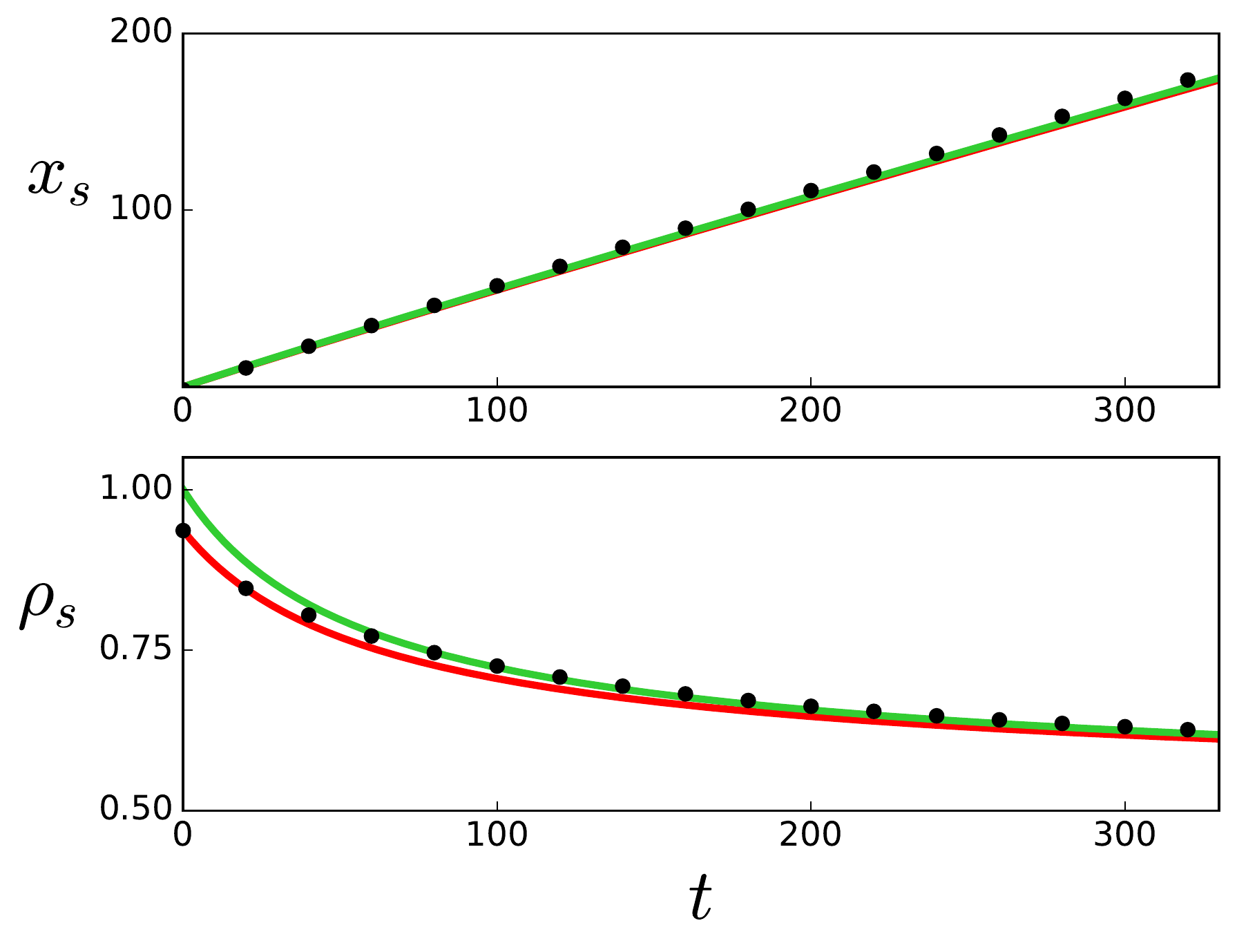}
  \caption{Position $x_s(t)$ and density $\rho_s(t)$ of the solitonic
    edge of the DSW induced by the initial profile
    \eqref{InitialFoeDSWPos} and \eqref{u_init}. The black dots are
    obtained by numerical integration of Eq.~\eqref{gNLS}. The colored
    solid lines are obtained from the analytic formulas \eqref{tParamLP}.
    In each plot the green theoretical solid line
    is obtained by taking $\rho_m=1$ and the red one by taking
    $\rho_m=0.9362$ (see the text).}
  \label{fig:seven}
\end{figure}
Eqs.~(\ref{tParamLP}) give the parametric dependence of the soliton
edge position $x_s(t)$ and density $\rho_s(t)$, which are represented
in Fig.~\ref{fig:seven} by continuous colored curves. For the analytic
determination of $\rho_s(t)$ and $x_s(t)$ we used in Eqs.~(\ref{tParamLP})
two possible
values of $\rho_m$: 1 (green curve in Fig.~\ref{fig:seven}) and
0.9362 (red curve). As one can see, there is no much difference,
and the theoretical results both agree well with the numerical
simulations (black dots). Of course, the results at short time for
$\rho_s$ are better when one takes the initial $\rho_m=0.9362$, but
this has little incidence at large time, and the results for $x_s$
almost do not depend on the choice $\rho_m=1$ or 0.9362.

For our initial parameters, the analytic formula \eqref{v_small_amp}
gives the asymptotic propagation velocity of the right small-amplitude
edge of the DSW: $dx_r/dt=1.11$. The numerical value of the velocity
is $1.12$ for our choice of initial condition, an agreement which
should be considered as very good since the position of the
small-amplitude edge is not easily determined, as is clear from
Fig.~\ref{fig:six}, and we evaluate it by means of an approximate
extrapolation of the envelopes of the wave at this edge.

\subsubsection{The parabolic initial condition
  \eqref{InitialParabola}}\label{sec412}

We now consider the initial density profile \eqref{InitialParabola}
with $u(x,0)=0$; that is, an initial state which is not of simple wave
type, and for which the formulas of the previous Sec. \ref{sec411} are
not directly applicable. In this case, the wave breaking of the right
propagating part of the pulse occurs in region $\rom{3}$, where $r_-$
is constant and equal to $-r_0$ at
$\rho_s(\twb)=\rho(x_s(\twb),\twb)=\rho_0$. Hence Eq. \eqref{9.4}
should be integrated with the boundary condition $\tal(\rho_0)=1$. In
a first stage of its development\footnote{It is explained below what
  happens in a second stage, see Eq.~\eqref{ip11}.}, the DSW matches
the dispersionless profile described by Eq.~\eqref{swrIII} where
$W^{(3)}(r_+,r_-)$ is given by Eq. \eqref{W13fin} and $\ox(r)$
by \eqref{ox-parabol}.

In the region of interest for us $r_-=-r_0$, so all quantities in
Eq.~\eqref{swrIII} depend on $r_+$ only. Differentiation with respect to $r_+$
yields, at the solitonic edge where $dx_s=V_s dt$:
\begin{equation}\label{ip1}
  (V_s-v_+)\frac{dt}{dr_+} -\frac{dv_+}{dr_+}\, t =
\frac{\partial^2W^{(3)}}{\partial r_+^2}
  \; ,
\end{equation}
where all quantities are evaluated at $r_s=r_+(x_s(t),t)$ and
$r_-=-r_0$. It follows from \eqref{crprm} that
$dr_s=c(\rho_s)d\rho_s/\rho_s$, yielding
\begin{equation}\label{ip2}
  \frac{dv_+}{dr_+}=1+\frac{\rho_s}{c(\rho_s)}\,\frac{dc}{d\rho_s}
=\frac{3+\gamma\rho_s}{2(1+\gamma\rho_s)}.
\end{equation}
For explicitly evaluating the right hand side of Eq.~\eqref{ip1}
it is convenient to write
$\mathscr{R}(r_1,r_2)=f(r_1)/f(r_2)$ in Eq.~\eqref{W13fin}
[see expressions \eqref{curlyR} and \eqref{curlyR_gamma}] which leads to
\begin{equation}\label{ip3}
  \frac{\partial W^{(3)}}{\partial r_+} =
-\frac{f'(r_+-r_-)}{f(r_+-r_-)} \,
W^{(3)}
+ \frac{\ox(r_+) f(2r_+)}{f(r_+-r_-)}\; ,
\end{equation}
and
\begin{equation}\label{ip4}
\begin{split}
& \frac{\partial^2 W^{(3)}}{\partial r_+^2} =
\left[2\, \frac{f'^2(r_+-r_-)}{f^2(r_+-r_-)}
-\frac{f''(r_+-r_-)}{f(r_+-r_-)}\right]
W^{(3)}\\
& - 2\, \ox(r_+)\,\frac{f'(r_+-r_-)f(2r_+)-f(r_+-r_-)f'(2r_+)}{f^2(r_+-r_-)}\\
& +\frac{d\ox}{dr_+} \frac{f(2r_+)}{f(r_+-r_-)}
\; .
\end{split}
\end{equation}
At the wave-breaking time $\twb$ one has
$r_s=r_0$; Eqs.~\eqref{W13fin}, \eqref{ip3}
and \eqref{ip4} then directly yield $W^{(3)}=0$,
$\partial W^{(3)}/\partial r_+=\ox(r_0)=l$ and [using
Eqs.~\eqref{ox-parabol} and \eqref{rhobder-parabol}]
\begin{equation}\label{ip5}
\frac{\partial^2 W^{(3)}}{\partial r_+^2}=
\left.\frac{d\ox}{dr}\right|_{r=r_0}=
-l \, \frac{(1+\gamma\rho_0)\sqrt{\rho_0}}{\rho_m-\rho_0} .
\end{equation}
At the wave breaking $V_s=v_+$ [$\tal(\rho_0)=1$ in \eqref{vsderhos}]
and Eq. \eqref{ip1} thus yields
\begin{equation}\label{ip6}
\twb=\frac{l}{\rho_m-\rho_0}
\frac{2\sqrt{\rho_0}(1+\gamma \rho_0)^2}{3+\gamma\rho_0}.
\end{equation}
This result is in agreement with the findings of Ref.~\cite{ikp-19c} and
 reduces to the one obtained in Ref.~\cite{ikp-19a} for
the nonlinear Schr\"odinger equation in the limit $\gamma=0$. For the
example considered in Fig.~\ref{fig:one}, expression \eqref{ip6}
gives $\twb=12.12$.

For times larger than $\twb$ one should solve Eq.~\eqref{ip1} using
the generic form \eqref{ip4} of its right hand side. It is convenient
to multiply this equation by $\rho_s/c(\rho_s)(d r_s/d\rho_s)=1$, which
yields
\begin{equation}\label{ip7}
  \rho_s\left(\tal(\rho_s)-1\right) \frac{dt}{d\rho_s}
-\frac{3+\gamma\rho_s}{2(1+\gamma\rho_s)}
\, t =
  \frac{\partial^2W^{(3)}(r_s,-r_0)}{\partial r_s^2}.
\end{equation}
The solution reads
\begin{equation}\label{ip8}
t(\rho_s)=\int_{\rho_0}^{\rho_s} \frac{\partial^2W^{(3)}}{\partial r_+^2}
\frac{G(\rho_s,\rho)}{\rho (\tal(\rho)-1)}\, d\rho,
\end{equation}
where the expression of $G$ is given in \eqref{GParam} and $r_+$ and
$\ox$ should be computed as functions of $\rho$ {\it via}
Eqs.~\eqref{eq17}, \eqref{rhobder-parabol} and \eqref{ox-parabol}.
Once $t(\rho_s)$ is
known, $x_s(\rho_s)$ is determined from Eq.~\eqref{swrIII}.
One can check that expression \eqref{ip8} yields
the correct result \eqref{ip6} for $\twb$:
When $\rho$ is close to $\rho_0$ one gets from Eq.~\eqref{GParam} and
\eqref{9.4} (with $\tal(\rho_0)=1$):
\begin{equation}
G(\rho_s,\rho)\simeq \left(\frac{\rho-\rho_0}{\rho_s-\rho_0}\right)^{3/2}\; .
\end{equation}
Inserting this expression back into \eqref{ip8} one obtains
\begin{equation}
t(\rho_0)=-\frac{2(1+\gamma\rho_0)}{3+\gamma\rho_0}
\left(\frac{\partial^2W^{(3)}}{\partial r_+^2}\right)_{r_+=r_0}\; .
\end{equation}
That is, as expected, $t(\rho_0)=\twb$, where $\twb$ is given by
Eq.~\eqref{ip6}.

The solution \eqref{ip8} is
acceptable only up to a time which we denote as $t_{\rm 2|3}$ at which
$r_s=r_m$, i.e., $\rho_s$ reaches a maximum which we denote as
$\rho_{\sss\rm M}$ where [see Eq.~\eqref{PhysVar_rho}]
\begin{equation}\label{rhoM}
\rho_{\sss\rm M} = \gamma^{-1}\tan^2[\sqrt{\gamma}(r_m+r_0)/2]\; .
\end{equation}
At times larger than $t_{2|3}$ the region $\rom{3}$ disappears and the
DSW is in contact with region $\rom{2}_r$ of the dispersionless
profile\footnote{The equivalent time was denoted as $t_{\rm A|B}$ in
  Ref.~\cite{ikp-19a}.} in which Eq.~\eqref{swrIIr} applies. Instead
of \eqref{ip7} one should thus solve
\begin{equation}\label{ip9}
  \rho_s\left(\tal(\rho_s)-1\right) \frac{dt}{d\rho_s}
-\frac{3+\gamma\rho_s}{2(1+\gamma\rho_s)}
\, t =
  \frac{\partial^2W^{(2)}(r_s,-r_0)}{\partial r_s^2},
\end{equation}
with the initial condition $t(\rho_{\sss\rm M})=t_{2|3}$.  $W^{(2)}$
in the above equation is given by Eq. \eqref{W2fin} which can be cast
in the form
\begin{equation}\label{ip10}
\begin{split}
W^{(2)}(r_+,r_-)= \frac{1}{f(r_+-r_-)}
& \Big[
\int_{-r_-}^{r_m} \ox(r) f(2r) dr \\
& + \int_{r_+}^{r_m}\ox(r) f(2r) dr
\Big].
\end{split}
\end{equation}
This yields
\begin{equation}\label{ip11}
\frac{\partial W^{(2)}}{\partial r_+}=- \frac{f'(r_+-r_-)}{f(r_+-r_-)} \,
W^{(2)}
-\frac{\ox(r_+) f(2r_+)}{f(r_+-r_-)},
\end{equation}
and
\begin{equation}\label{ip12}
\begin{split}
& \frac{\partial^2 W^{(2)}}{\partial r_+^2} =
\left[2\, \frac{f'^2(r_+-r_-)}{f^2(r_+-r_-)}
-\frac{f''(r_+-r_-)}{f(r_+-r_-)}\right]
W^{(2)}\\
& + 2\, \ox(r_+)\,\frac{f'(r_+-r_-)f(2r_+)-f(r_+-r_-)f'(2r_+)}{f^2(r_+-r_-)}\\
& -\frac{d\ox}{dr_+} \frac{f(2r_+)}{f(r_+-r_-)}.
\end{split}
\end{equation}
The solution of \eqref{ip9} reads
\begin{equation}\label{ip13}
t(\rho_s)=t_{2|3}+
\int_{\rho_{\sss\rm M}}^{\rho_s} \frac{\partial^2W^{(2)}}{\partial r_+^2}
\frac{G(\rho_s,\rho)}{\rho (\tal(\rho)-1)}\, d\rho.
\end{equation}
Once $t(\rho_s)$ is determined from \eqref{ip13} $x_s(\rho_s)$ is
obtained from \eqref{swrIIr}.
Formulas \eqref{ip8}, \eqref{swrIII}, \eqref{ip13} and \eqref{swrIIr}
give parametric expressions of $\rho_s(t)$ and $x_s(t)$ for all
$t\ge \twb$.

We note here that in the limit $\gamma\to 0$, the solution of
Eq. \eqref{9.4} reads $\tal=-1+2\sqrt{\rho_0/\rho}$ which, from
\eqref{vsderhos}, yields $V_s=\sqrt{\rho_s}=\tfrac12(r_s-r_-)$, in agreement
with the findings of Ref. \cite{ikp-19a}. Also in this limit, one
gets the exact expression $G(\rho_s,\rho) =
(\sqrt{\rho}-\sqrt{\rho_0})^{3/2}(\sqrt{\rho_s}-\sqrt{\rho_0})^{-3/2}$
and formulas \eqref{ip8} and \eqref{ip13} reduce to the equivalent
ones derived for the NLS equation in Ref. \cite{ikp-19a}.

\begin{figure}[t!]
  \centering
  \includegraphics[width=\linewidth]{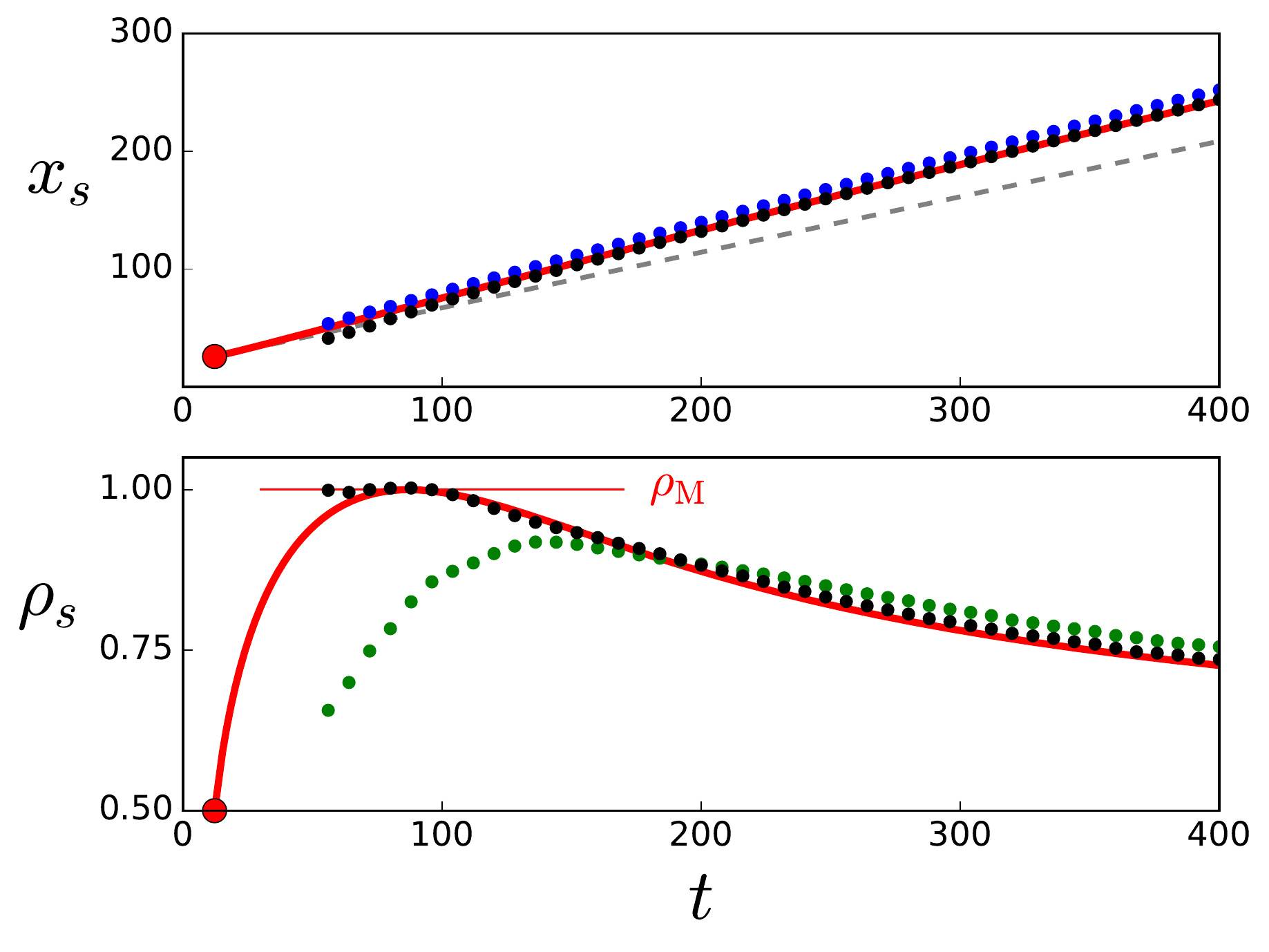}
  \caption{Position $x_s(t)$ and density $\rho_s(t)$ of the solitonic
    edge of the DSW induced by the initial profile
    \eqref{InitialParabola} with $\rho_0=0.5$, $\rho_m=2$ and
    $l=20$. The continuous red curves are the theoretical results. The
    big red dots locate the wave breaking time $\twb=12.12$, position
    $x_s(\twb)=25.71$ and intensity $\rho(x_s(\twb),\twb)=\rho_0$. The
    dotted curves are extracted from numerical simulations (see the
    text). The dashed gray line in the upper panel corresponds to a
    soliton edge which would propagate at the background sound
    velocity. The horizontal red line in the lower panel marks the
    position of the maximum $\rho_{\sss\rm M}$ as given by
    Eq.~\eqref{rhoM}.}
  \label{fig:soliton_edge2}
\end{figure}

In the generic case $\gamma\ne 0$, the results for $x_s(t)$ and
$\rho_s(t)$ corresponding to the parameters of Fig.~\ref{fig:one} are
compared in Fig.~\ref{fig:soliton_edge2} with the values extracted
from the numerical solution of Eq.~\eqref{gNLS} (with $\gamma=1$).
The position $x_s$ of the solitonic edge is easily determined from the
numerical simulation: it is located between the maximum of the smooth
part of the intensity and the following zero of $\rho-\rho_0$
(respectively black and blues dots in the upper panel of the figure).
The intensity $\rho_s$ of the solitonic edge is more difficult to
extract from the numerical simulations in the present situation that
in the case studied in Fig.~\ref{fig:seven}. At times close to the
theoretical wave-breaking time, one cannot exactly decide from the
numerical intensity profile if the oscillations at the boundary of
the right propagating edge of the region of increased intensity are
linear disturbances (due to dispersive effects) or correspond to the
birth of a DSW. It is only after a certain amount of time that the
oscillations become clearly nonlinear. Even then, the precise location
of the intensity of the solitonic edge is not easily determined: it is
comprised between the maximum intensity of the smooth part of the
spectrum and the following maximum intensity (respectively black and
green dots in the lower panel of Fig.~\ref{fig:soliton_edge2}). The
green dots initially yield a clear under estimate of $\rho_s$, and, on
the contrary, when the dispersive shock if fully developed, they
indicate a too large value. This is due to the fact that, at large
time, the amplitude of the oscillations in the vicinity of the soliton
edge increase within the DSW.

We note also that expression \eqref{rhoM} gives a simple and accurate
prediction for the maximum of $\rho_s$ ($\rho_{\sss\rm M}=1$ in the
case considered in Fig.~\ref{fig:soliton_edge2}). At times large
compare to the time $t_{2|3}$ at which this maximum is reached, one can
approximately evaluate the behavior of $\rho_s$ and $x_s$ as follows:
The integrands in Eqs. \eqref{ip8} and \eqref{ip13} are of the form
$(\rho_s-\rho_0)^{-3/2}F^{(i)}(\rho_s,\rho)$ where $F$ is a non
singular function [with $i=2$ for Eq. \eqref{ip8} and $i=3$ for
Eq. \eqref{ip13}]. Since at large time $\rho_s$ tends to $\rho_0$,
formula \eqref{ip13} can be approximated by
\begin{equation}
t(\rho_s)\simeq \frac{{\cal A}}{(\rho_s-\rho_0)^{3/2}},
\end{equation}
where
\begin{equation}
{\cal A}=\int_{\rho_0}^{\rho_{\sss\rm M}}
\left(F^{(3)}(\rho_0,\rho)-F^{(2)}(\rho_0,\rho)\right)d\rho .
\end{equation}
From there it follows that, at large $t$, $\rho_s$ tends to $\rho_0$
as $t^{-2/3}$ and $x_s$ exceeds $c(\rho_0) t$ by a factor of
order $t^{1/3}$. The fact that the soliton edge propagates at a
velocity higher that the speed of sound is illustrated in
Fig. \ref{fig:soliton_edge2} by the difference between $x_s(t)$ and
the gray dashed line of equation $x=x_s(\twb)+c(\rho_0)(t-\twb)$.

A final remark is in order here. When $\rho_m$ gets large compared
with $\rho_0$, $\tal(\rho_s)$---solution of \eqref{9.4}---may become
negative. For instance, when $\gamma=0$ this occurs for
$\rho_m\ge 4\,\rho_0$. As noticed in Ref.~\cite{egkkk-07}, this is
linked to the occurrence within the DSW of a vacuum point \cite{El1995} at
which the density cancels. In this case, one should use another branch
in the dispersion relations \eqref{linlaw} and \eqref{omegatild}, but
once this modification is performed, the approach remains perfectly
valid and accurate.

  When $\gamma>0$, a more serious problem occurs at larger values of
  $\rho_m$, when $\tal(\rho_s)$ reaches $-\tfrac12$.  A rough estimate
  based on the regime $\gamma\ll 1$ indicates that this occurs
  starting from $\rho_m\approx 16\, \rho_0$. In this case the
  theoretical approach leads to singularities [see, e.g.,
  Eq.~\eqref{9.4}] whereas the numerical simulations do not indicate a
  drastic change of behavior in the dynamics of the DSW. This points
  to a probable failure of the Gurevich-Meshcherkin-El approach,
  however the study of this problem is beyond the scope of the present
  work and we confine ourselves to a regime of initial parameters for
  which $\tal(\rho_s)$ remains larger than $-\tfrac12$ and the theory
  is applicable.

\subsection{Negative pulse}\label{sec42}

In this section we consider an initial condition with a region where
the density is depleted with respect to that of the background. For
simplicity we only consider the case where, as in
Sec.~\ref{sec411}, the initial Riemann invariant $r_-(x,0)$ is
constant and equal to $-r_0$ for all $x$.

\begin{figure}[t!]
	\centering
	\includegraphics[width=\linewidth]{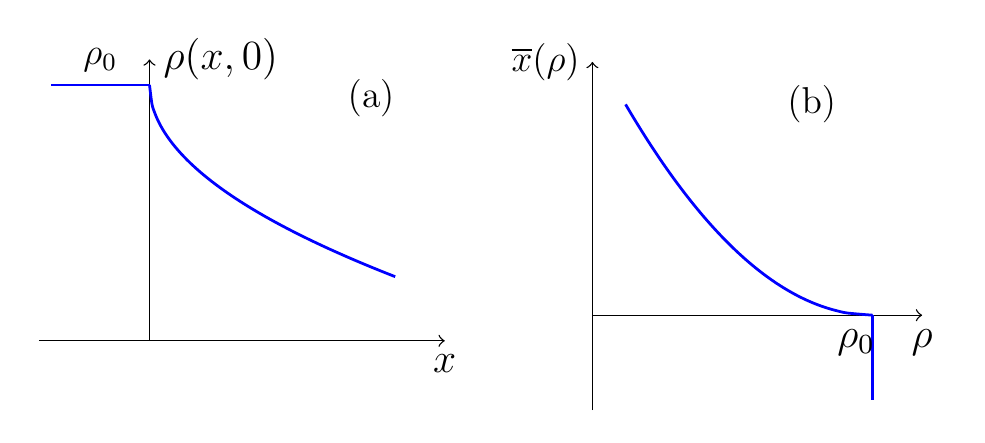}
	\caption{(a) Initial profile $\orho(x)$ of a monotonous
          negative pulse.  (b) Inverse function $\overline{x}(\rho)$.}
	\label{fig:eight}
\end{figure}

Let us first consider a monotonous initial pulse
\begin{equation}\label{8.23}
\rho(x,0)=
\begin{cases}
\rho_0 & \mbox{if}\quad x<0, \\
\rho_0-\widetilde{\rho}(x) & \mbox{if}\quad x\ge 0.
\end{cases}
\end{equation}
This initial condition is sketched in Fig.~\ref{fig:eight}.
The small-amplitude edge propagates with the group velocity
\begin{equation}\label{}
V_r(\rho_r)=\frac{d\omega}{dk}=u(\rho_r)+
\frac{\sqrt{\rho_r}}{1+\gamma\rho_r}\frac{2\al^2(\rho_r)-1}{\al(\rho_r)},
\end{equation}
where $\rho_r(t)$ is the density of the
small amplitude edge. Following a reasoning analogous to the one
already employed in Sec.~\ref{sec411} for the solitonic edge, we
parameterize all relevant quantities at the small amplitude edge in
term of $\rho_r$. This leads to the following differential equation
\begin{equation}\label{eqdiff-t-rhor}
\frac{\sqrt{\rho_r}}{1+\gamma\rho_r} \frac{2\al^2-\al-1}{\al}
\frac{d t}{d \rho_r} -
\frac{3+\gamma \rho_r}{2\sqrt{\rho_r}(1+\gamma\rho_r)^2}t =
\frac{d\ox(\rho_r)}{d \rho_r},
\end{equation}
which should be solved with the initial condition $\rho_r=\rho_0$ at
$t=0$. The
solution reads
\begin{equation} \label{tParamN}
t(\rho_r)=\int_{\rho_0}^{\rho_r}
{\frac{\al(\rho)(1+\gamma\rho)\ox'(\rho)G(\rho_r,\rho)}
{\sqrt{\rho}(\al(\rho)-1)(1+2\al(\rho))}}d\rho,
\end{equation}
where $\ox'(\rho)=d\ox/d\rho$ and
\begin{equation} \label{}
\begin{split}
G & (\rho_r,\rho)= \\ &
\exp{\left(\int_{\rho}^{\rho_r}\!\!\frac{\al(\rho') (3+\gamma\rho')\, d\rho'}
{2\rho'(1+\gamma\rho')(\al(\rho')-1)(1+2\al(\rho'))}\right)}.
\end{split}
\end{equation}
Consequently, the position $x_r$ of the small amplitude edge is given by
\begin{equation}\label{xParamN}
x_r(\rho_r) =\left[u(\rho_r) + c(\rho_r)\right]t(\rho_r)+\ox(\rho_r),
\end{equation}
where $u(\rho)$ is given by Eq. \eqref{eq17}. Formulas \eqref{tParamN}
and \eqref{xParamN} determine, in a parametric form, the coordinates
$x_r(t)$ and $\rho_r(t)$ of the small-amplitude edge.

\begin{figure}[t!]
\centering
\includegraphics[width=\linewidth]{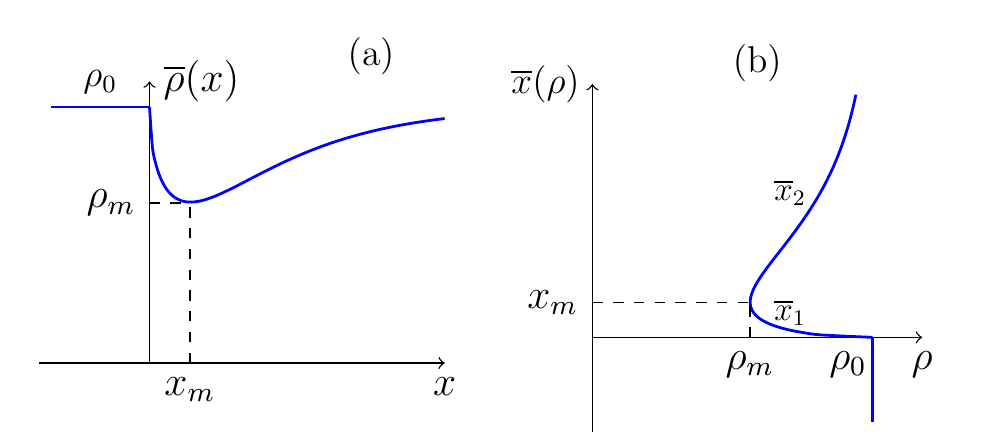}
\caption{(a) Initial profile $\orho(x)$ of a non-monotonous
          negative pulse.  (b) The inverse function
          $\overline{x}(\rho)$ is represented by two branches
         $\overline{x}_1(\rho)$ and $\overline{x}_2(\rho)$. Here
          $x_m$ is the point where $\rho$ takes its minimal value
          $\rho_m$.}
\label{fig:nine}
\end{figure}

In the case of a non-monotonous negative pulse, such as the one
represented in Fig. \ref{fig:nine}, the approach has to be modified in
a manner similar to that exposed in Sec. \ref{sec411} for the
non-monotonous profile displayed in Fig. \ref{fig:five}. We do not
write down the corresponding formulas for not overloading the paper
with almost identical expressions. We only indicate that $\rho_m$
denotes now the minimal value of the density in the initial
distribution.

\begin{figure}[t]
\centering
\includegraphics[width=\linewidth]{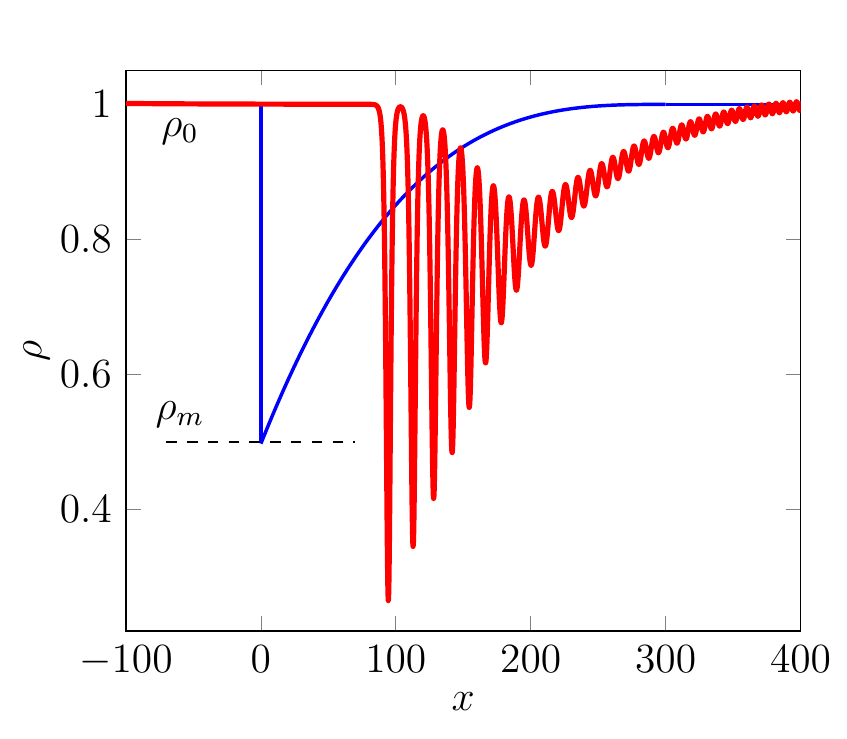}
\caption{Density profile $\rho(x,t=300)$ (red solid curve), which
  results from the time evolution of the simple-wave initial condition
  (\ref{InitialFoeDSWNeg}) with $\rho_0=1$, $\rho_m=0.5$ and $x_0=300$
  (blue solid curve).}\label{fig:ten}
\end{figure}

For numerically testing our approach we consider
a non-monotonous initial profile of the form
\begin{equation}\label{InitialFoeDSWNeg}
\orho(x) =
\begin{cases}
\rho_0 -(\rho_m-\rho_0) \left(\frac{x}{x_0}-1\right)^3
\; \mbox{if}\; x\in[0,x_0],\\
\rho_0 \quad\mbox{elsewhere},
\end{cases}
\end{equation}
see the blue
curve in Fig.~\ref{fig:ten}. In this case one has
$\overline{x}_1(\rho)=0$ and 
\begin{equation}\label{}
\overline{x}_2(\rho\le \rho_0) = x_0-
x_0\left( \frac{\rho_0-\rho}{\rho_0-\rho_m} \right)^{1/3}.
\end{equation}
The initial velocity distribution $u(x,0)$ is determined from
\eqref{InitialFoeDSWNeg} by imposing that $r_-(x,0)=-r_0$. The typical
wave pattern for the light evolved from such a pulse is shown in
Fig.~\ref{fig:ten}.

The motion of the right, small amplitude edge is determined as
previously explained. Eq. \eqref{9.5} should be solved with the
initial condition $\alpha(\rho_m)=1$. Then, in the non-monotonous case
we consider here, Eq. \eqref{eqdiff-t-rhor} should be solved with an
initial condition different from the one used in the case of the
initial profile \eqref{8.23}. The appropriate initial value is here
$\rho_r=\rho_m$ at $t=0$.  As a result, the lower boundary of
the integration region in \eqref{tParamN} should be changed from
$\rho_0$ to $\rho_m$.
\begin{figure}[t]
\centering
\includegraphics[width=\linewidth]{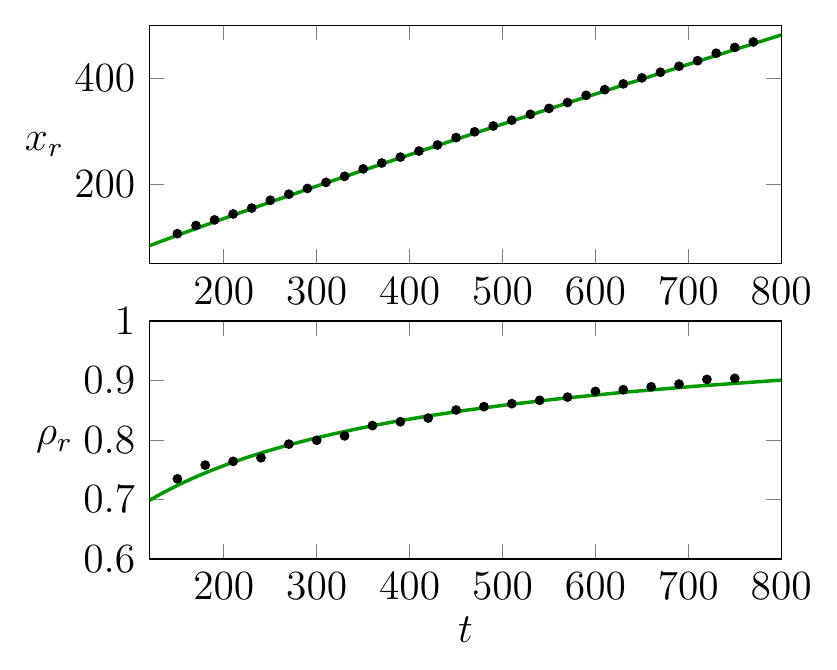}
\caption{Coordinate $x_r(t)$ and density $\rho_r(t)$ of the
  small-amplitude edge of the DSW induced by the initial profile
  \eqref{InitialFoeDSWNeg}. The black dots are obtained by numerical
  integration of Eq.~\eqref{gNLS}. The green solid lines are obtained
  from the analytic solution, i.e., from Eqs. \eqref{tParamN} and
  \eqref{xParamN}.}\label{fig:eleven}
\end{figure}
The theoretical results compare very well with the one extracted from
the numerical simulations, as illustrated in Fig.~\ref{fig:eleven}.
Note that the determination of the small amplitude edge of the
numerical DSW is a delicate task, which is simplified if the DSW
contains a large number of oscillations. This is the reason why we had
to chose here an initial profile of extension larger than the one
considered in the previous section \ref{sec411} ($x_0=300$ instead of
50).  Note however that the case
considered here presents an advantage compared to the similar one
considered in Sec.~\ref{sec411}: the motion of the small-amplitude
edge is less dependent of the precise characteristics of $\orho(x)$
at its extremum than the motion of the solitonic edge. As a result, we
need not discuss here the precise numerical implementation of the
idealized discontinuous profile \eqref{InitialFoeDSWNeg}, contrarily
with what has been done in Sec.~\ref{sec411}.

At asymptotically large time the velocity of the trailing soliton
generated from a localized pulse is determined from the knowledge of
$\tal(\rho_0)$, where $\tal(\rho)$ is the solution of equation
(\ref{9.4}) with the boundary condition $\tal(\rho_m) = 1$.  Within
this approximation, the asymptotic velocity of the soliton edge reads
\begin{equation}\label{SolVelN}
V_s(t\to\infty)=\frac{\sqrt{\rho_0}}{1+\gamma\rho_0}\tal(\rho_0).
\end{equation}

\begin{figure}[t]
\centering
\includegraphics[width=\linewidth]{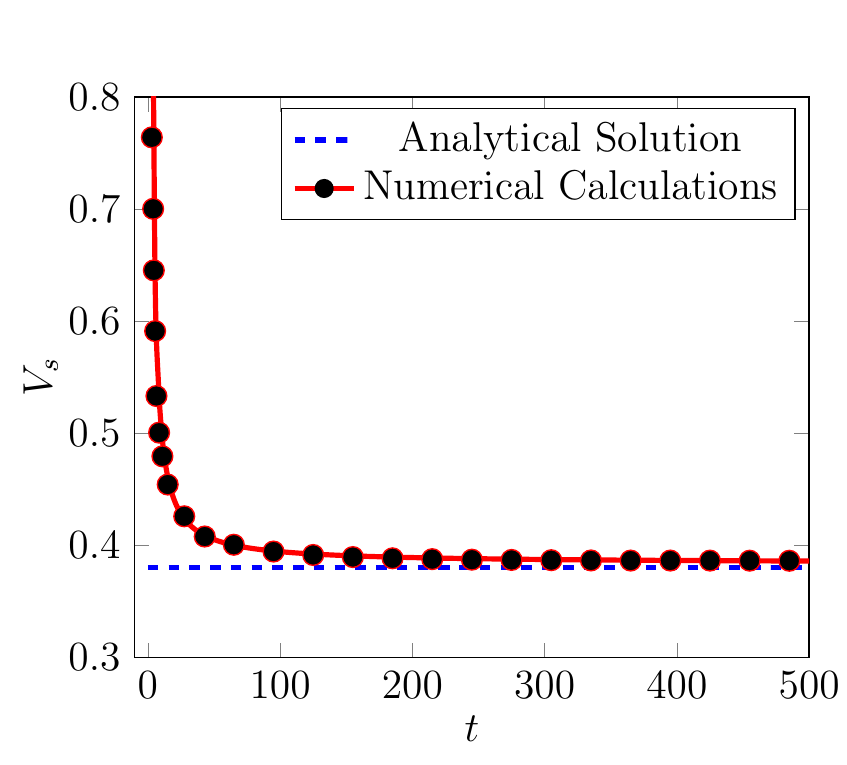}
\caption{Black dots: numerically determined velocity $V_s(t)$ of the
  soliton edge.  The initial conditions are specified in
  Eq.~(\ref{InitialFoeDSWNeg}) with $\rho_0=1$, $\rho_m=0.7$ and
  $x_0=300$ (blue). Continuous red line: fit of the numerical data by
  the formula: $V_s(t)=V_s^{\rm num}+\sum_{i=1}^6 a_i t^{-i}$.  One
  obtains $V_s^{\rm num} = 0.39$, in good agreement with the
  theoretical prediction from Eq.~(\ref{SolVelN}): $V_s(t\to\infty) =
  0.37$ (blue dashed line).} \label{fig:twelve}
\end{figure}

The comparison of the analytic prediction (\ref{SolVelN}) with our
numerical simulations is not difficult because the soliton edge is
easily identified in the numerical density profiles, and, indeed, a
leading soliton is easily located at this edge of the numerically
determined DSW. As we see in Fig.~\ref{fig:twelve}, its velocity tends
at large $t$ to a constant value. In this figure, the numerical result
is fitted with the empirical formula $V_s(t) = V_s^{\rm
  num}+\sum_{i=1}^6 a_i t^{-i}$, where $V_s^{\rm num}$ and $a_i$ are
fitting parameters. The trend is in excellent agreement with the
prediction (\ref{SolVelN}) since one obtains $V_s^{\rm num} = 0.39$
whereas from (\ref{SolVelN}) one expects $V_s(t\to\infty) = 0.37$.

\section{Conclusion} \label{sec5}

In this work we have presented a theoretical study of the dynamics of
spreading of a pulse of increased (or decreased) intensity propagating
over a uniform background. The initial non-dispersive stage of evolution is
described by means of Riemann's method, the result of which compares
very well with numerical simulations. After the wave breaking time, we
have studied the behavior of the resulting dispersive shock wave at
its edges, by mean of the modification of El's method presented in
Ref.~\cite{Kamchatnov-19}. Here also the results compare very well
with the ones of numerical simulations.

Our work represents a comprehensive theoretical description of the
spreading, of the wave breaking and of the subsequent formation of a
dispersive shock in a realistic setting for a system described by a
nonintegrable nonlinear equation. In particular, our approach yields a
simple analytic expression for the wave-breaking time, even in
situations where the initial density and velocity profiles do not
reduce to a simple wave configuration. Also, in view of future
experimental studies, we have devoted a special attention to the
determination of the position and intensity of the solitonic edge of
the DSW issued from the spreading of a region of increased light
intensity.

\begin{acknowledgments}
  We thank T. Bienaim\'e and M. Bellec for fruitful discussions.
  J.-E.S thanks Moscow Institute of Physics and Technology
  and Institute of Spectroscopy of Russian Academy of Sciences for
  kind hospitality during his internship there, when this research was
  started. S.K.I and A.M.K thank RFBR for financial support of this study
  in framework of the project 20-01-00063.
\end{acknowledgments}

\end{document}